\documentclass{jfm}

\usepackage{graphicx}
\usepackage{newtxtext}
\usepackage{newtxmath}
\usepackage{natbib}
\usepackage{hyperref}
\usepackage{anyfontsize}
\hypersetup{
    colorlinks = true,
    urlcolor   = blue,
    citecolor  = blue,
}

\newcommand{\RomanNumeralCaps}[1]
\linenumbers

\title{Measured multiple flow states in turbulent thermal convection with aspect ratio 10}

\author{Yi-Zhen Li\aff{1,2},
Jun-Jie Huo\aff{1},
Xin Chen\aff{1}\corresp{\email{xinchen99@nwpu.edu.cn, hengdongxi@nwpu.edu.cn}}
\and
Heng-Dong Xi\aff{1}\footnotemark[1]
}

\affiliation{
\aff{1}
Institute of Extreme Mechanics,
School of Aeronautics,
National Key Laboratory of Aircraft Configuration Design and Key Laboratory for Extreme Mechanics of Aircraft of Ministry of Industry and Information Technology,
Northwestern Polytechnical University, Xi'an, 710072, Shannxi, China
\aff{2}
Taihang national laboratory, Chengdu, 610000, Sichuan, China
}

\begin{document}
\maketitle

\begin{abstract}
We report an experimental investigation of turbulent Rayleigh–Bénard convection in a rectangular cell of large aspect ratio ($\Gamma = 10$) over the Rayleigh number range $5.4\times10^7 \le Ra \le 7.2\times10^9$ and Prandtl number range $4.3 \le Pr \le 67.3$.
Planar particle image velocimetry measurements show that the flow self-organises into several horizontally stacked convection rolls, and repeated experiments under identical parameters (both $Ra$ and $Pr$) reveal that the number of rolls varies within the range of 3 to 7 with 6 being the most probable, which demonstrates the presence of multiple flow states.
When $Pr$ is increased to 67.3, the number of roll-like structures increases significantly, indicating a structural transition from a roll-dominated to a plume-dominated flow.
This transition is reflected in the global momentum transport, for $Pr \leq 18.3$ the Reynolds number scales as $Re \sim Ra^{0.58}Pr^{-0.97}$, whereas the scaling is changed to $Re \sim Ra^{0.72}$ when $Pr$ reaches 67.3.
Within individual rolls, we further examine the Reynolds numbers based on horizontal and vertical velocity components, $Re_{u,\text{roll}}$ and $Re_{w,\text{roll}}$, and find that the former increases while the latter decreases with roll size (quantified as the aspect ratio of the roll $\Gamma_\text{roll}$) due to continuity constraints, with their ratio following $Re_{w,\text{roll}}/Re_{u,\text{roll}} \sim \Gamma_\text{roll}^{-0.61}$.
We impose different initial flow conditions (roll structures) with controlled perturbations, and demonstrate that the initial condition can influence the final turbulent state.
We show that the number of horizontally stacked rolls regulates the global transport: larger number of rolls induces greater vertical momentum and heat transfer.
Our study provides the first systematic experimental evidence of multiple flow states in large aspect ratio turbulent Rayleigh–Bénard convection and clarify how these states influence global transport.
\end{abstract}


%{\bf MSC Codes }  {\it(Optional)} Please enter your MSC Codes here

\section{Introduction}
\label{sec:intro}
Multiple flow states reveal the nonlinear, multistable nature of turbulence within both natural and laboratory systems.
In geophysical and astrophysical flows, large-scale turbulent flows can adopt distinct quasi stable configurations even when the external conditions remain nearly identical.
Notable examples include the bimodal path of the Kuroshio Current in the western Pacific \citep{schmeits2001bimodal}, the alternation between blocking and zonal regimes in mid-latitude atmospheric flows \citep{rahmstorf2002ocean}, and the multiple zonal jet structures observed on giant planets such as Jupiter \citep{bouchet2019rare}.
These examples show that even at high Reynolds numbers, where turbulence is expected to be fully developed, the organization of the large scale flow can remain path dependent and non-unique.
Such non-uniqueness can result in substantial differences in global transport properties including heat, momentum, and mass fluxes.

In recent years, both laboratory experiments and numerical simulations have increasingly reported comparable manifestations of multiple flow states under well-controlled boundary conditions.
Canonical systems exhibiting this phenomenon include Taylor–Couette flow \citep{Huisman2014NC, vanderVeen2016PRF, Xia2017JFM}, planar Couette flow with spanwise rotation \citep{xia2018multiple}, spherical Couette flow \citep{zimmerman2011bi}, von Kármán swirling flow \citep{Ravelet2004PRL}, two-dimensional shear flow \citep{Dallas2020PRF}, double-diffusive convection \citep{yang2020multiple}, and channel flow \citep{Markeviciute2021JFM}.
Multiple flow states have also been identified in turbulent Rayleigh–Bénard (RB) convection \citep{Xi2008POF, vanderPoel2011PRE, Weiss2011JFM, Xie2018PRL, Wang2018PRF, Wang2020JFM, wang2020PRL, Zwirner2020PRL}.

Turbulent RB convection has drawn significant attention over the past decade not only due to its relevance in understanding thermally driven turbulence but also because of its implications for astrophysical and geophysical flows.
Representative examples include convection in oceans, planetary atmospheres, and the Earth's mantle \citep{Ahlers2009RMP, Lohse2010ARFM, Xia2013TAML, lohse2024RMP}.
A typical RB system consists of a fluid layer confined between a heated bottom plate and a cooled top plate.
The key control parameters include the Rayleigh number ($Ra=\alpha g \Delta T H^3/\nu \kappa$), which quantifies the strength of buoyancy forcing, the Prandtl number ($Pr=\nu / \kappa$), which characterises the fluid properties, and the aspect ratio ($\Gamma=L/H$), which describes the geometry of the convection cell.
Here, $\alpha$, $\nu$, and $\kappa$ are the thermal expansion coefficient, kinematic viscosity, and thermal diffusivity of the working fluid, respectively.
The quantity $g$ denotes the gravitational acceleration, while $\Delta T$ is the imposed temperature difference between the plates.
The dimensions $L$ and $H$ represent the horizontal length and vertical height of the cell.

With respect to the existence of multiple flow states in turbulent RB convection, \citet{Xi2008POF} first reported an experimental observation in a $\Gamma = 0.5$ RB cell, where the flow randomly switches between a single-roll structure and a vertically stacked double-roll structure.
They further demonstrated that the single-roll structure enhanced heat transport relative to the vertically stacked double-roll configuration.
This form of multistability, involving transitions between distinct flow states under identical control parameters, was subsequently confirmed through both numerical simulations and additional experiments \citep{Ahlers2009NJP, vanderPoel2011PRE, Weiss2011JFM}.
However, these studies were primarily restricted to RB systems with aspect ratio $\Gamma$ smaller than unity.
By contrast, multiple flow states in natural systems such as the ocean and atmosphere arise in geometries with much larger aspect ratios.
For large $\Gamma$ turbulent RB system, previous work \citep{wang2020PRL,Wang2018PRF,xu2023long} showed horizontally stacked multi roll configurations, and the number of rolls depends strongly on the initial conditions and on the boundary conditions.
Specifically, \citet{wang2020PRL} studied a large $\Gamma$ system under horizontally periodic boundary conditions and found that distinct turbulent states can be triggered by changes in the initial conditions.
The range of allowable roll numbers was further elucidated with the aid of elliptic instability theory \citep{wang2020PRL, Shishkina2021PRF}.
In addition, \citet{Wang2018PRF} identified the presence of multiple flow states in a fully confined system with no-slip and impermeable boundary conditions at all walls.
They provided a phase diagram that relates the number of turbulent states to both the aspect ratio and the tilt angle of the convection cell.
Despite these advances, experimental evidence for multiple flow states in large aspect ratio RB convection remains absent.
Further gaps arise because the range of roll numbers observed in simulations is broader than that reported in experiments.
For example, \citet{Xia2008conference} observed a triple roll structure in a quasi two dimensional system with $\Gamma = 9.9$, which is lower than the minimum roll number reported in simulations \citep{Wang2018PRF}.
Three dimensional effects may account for part of this discrepancy, yet direct experimental validation is still lacking.
A systematic experimental investigation is therefore needed.

In this work, we carry out a systematic investigation of flow structures in turbulent RB convection with $\Gamma=10$ over a wide parameter range and we assess their impact on momentum transfer and heat transfer.
The remainder of the paper is organized as follows.
\autoref{sec:setup} describes the experimental set-up and the measurement methods.
\autoref{sec:results} presents the main results, including typical flow structures, the manifestation of multiple flow states, the effect of roll size on momentum transfer, and the evolution of heat and momentum with the number of rolls.
\autoref{sec:conclusions} provides a brief summary and discusses implications for future studies.

\section{Experimental Set-up}\label{sec:setup}

\begin{figure}
\centerline{\includegraphics[width=0.75\columnwidth]{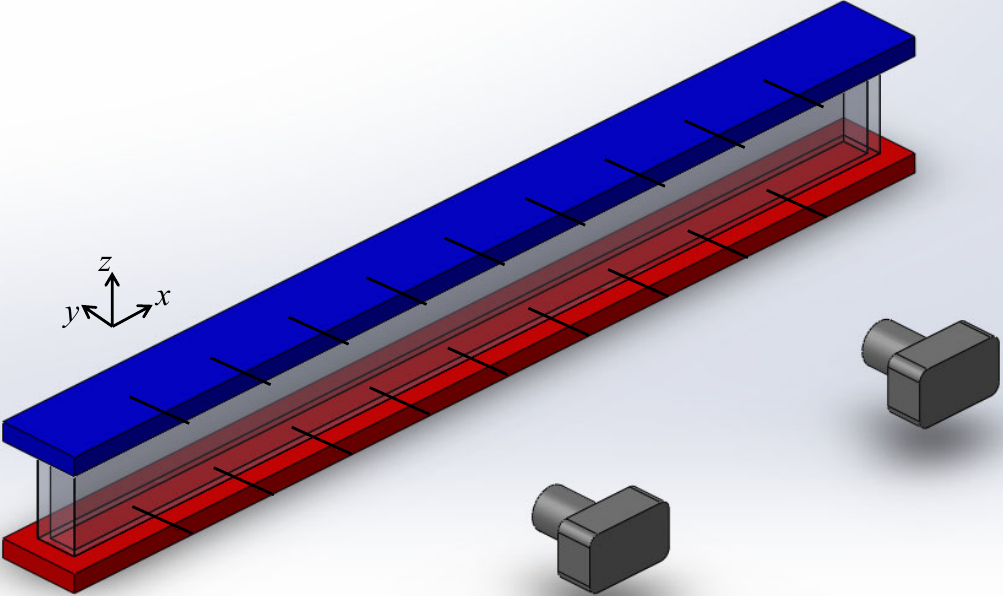}}
\caption{Schematic of the convection cell.
The blue and red plates represent the top and bottom plates, respectively.
Two cameras are arranged in a parallel configuration and are used to measure the flow field.}
\label{fig:f1}
\end{figure}

The convection system, as shown in \autoref{fig:f1}, consists of copper top and bottom plates and a Plexiglas sidewall.
The convection cell has a length ($L$, in x direction), width ($W$, in y direction), and height ($H$, in z direction) of $100\ \text{cm}$, $3\ \text{cm}$, and $10\ \text{cm}$, respectively.
Accordingly, the aspect ratios are $\Gamma = L/H = 10$ and $\Gamma_{lateral} = W/H = 0.3$.
The temperature of the top plate is regulated by a refrigerated circulator (Polyscience PP15R-40-A12Y), which continuously extracts heat through embedded channels within the plate.
The bottom plate is uniformly heated by electric resistance wires (Cr20Ni80) wrapped in glass-fibre insulation.
A long-term stable DC power supply (EA-PS 91500-30) delivers constant and spatially uniform heating power to the bottom plate.
Eighteen thermistors, each $2.5\ \text{mm}$ in diameter (nine on each plate), monitor the temperatures of the top and bottom plates.
These thermistors are evenly distributed along the length and positioned at the mid-width, as shown in \autoref{fig:f1}.
During the experiments, both plates were thermally insulated with foam inside a temperature-controlled room.
Throughout the experimental temperature range of $10^\circ C$ to $55^\circ C$, the maximum temperature variation across the top (bottom) plate was less than $0.15\%$ ($0.25\%$) of the mean plate temperature.
This confirms that both plates maintained excellent temperature uniformity during all measurements.
The flow structure in the present experiments is sensitive to the tilt of the cell.
Unless stated otherwise, all results reported here were obtained after careful levelling of the convection cell.

The velocity field was measured in the centre plane using planar Particle Image Velocimetry (PIV).
To achieve a wide field of view, the optical path was optimized following \citet{Lee2023design}, and the light-sheet thickness was maintained below $2\ \text{mm}$ throughout the measurement region.
Two cameras, each with a resolution of $3388\ \mathrm{pixels} \times 2712\ \mathrm{pixels}$, were positioned parallel to each other to capture simultaneous flow images, as shown in \autoref{fig:f1}.
The measurement regions of the two cameras were designed to overlap to enable seamless image reconstruction.
The full-field images were obtained by stitching the two half-field images based on calibrated features within the overlapping zone.
The velocity field was calculated using an interrogation windows of $32 \times 32$ pixels with a $50\%$ overlap.
The above-mentioned configuration yields approximately 32 velocity vectors in the vertical direction and 320 in the streamwise direction.
Polyamide seeding particles with a mean diameter of $20\ \mu m$ and a density of $1.03$–$1.05\ g/cm^3$ were used for flow tracing.
Except for three long duration measurements of at least $10$ hours that were sampled at $0.25\ \text{Hz}$, all other measurement sets lasted at least two hours with a sampling frequency of $1\ \text{Hz}$.
The longest continuous measurement spanned 22.2 hours.

To extend the experimental parameter space, different working fluids were employed, as listed in \autoref{table_fluid}.
The range of $Pr$ was mainly varied by adjusting the volume ratio of glycerol water mixtures.
For example, the Prandtl number of 9.9 corresponds to a glycerol water mixture with a volume ratio of 1:9, which is subsequently referred to as 0.1 glycerol.
The increase in $Ra$ was mainly achieved by using $0.65\ cSt$ silicone oil as the working fluid.

\begin{table}
\begin{center}
\def~{\hphantom{0}}
\begin{tabular}{cccccc}
Fluid         & $Pr$    & $Ra$ ($\times 10^8$) & $\alpha$ ($\times 10^{-4}$) & $\kappa$ ($\times 10^{-7}$) & $\nu$ ($\times 10^{-7}$)\\[3pt]
water         & 4.3  & $1.61\sim7.81$ & $3.85$   & $1.52$ & $6.58$    \\
water         & 6.1  & $0.59\sim3.92$ & $2.59$   & $1.46$ & $8.90$    \\
0.1 glycerol  & 9.9  & $0.83\sim3.51$ & $2.71$   & $1.34$ & $13.2$    \\
0.25 glycerol & 18.3 & $1.09\sim3.24$ & $3.76$   & $1.24$ & $22.6$    \\
0.5 glycerol  & 67.3 & $0.54\sim1.63$ & $5.23$   & $1.09$ & $73.3$    \\
silicon oil   & 9.9  & $31.7\sim72.0$ & $14.0$   & $0.66$ & $6.50$    \\
\end{tabular}
\caption{Working fluids and parameters of the experiments.}
\label{table_fluid}
\end{center}
\end{table}

\section{Results and discussions}\label{sec:results}
\subsection{Typical flow structure}\label{sec:Spontaneous}

\begin{figure}
\centerline{\includegraphics[width=1\columnwidth]{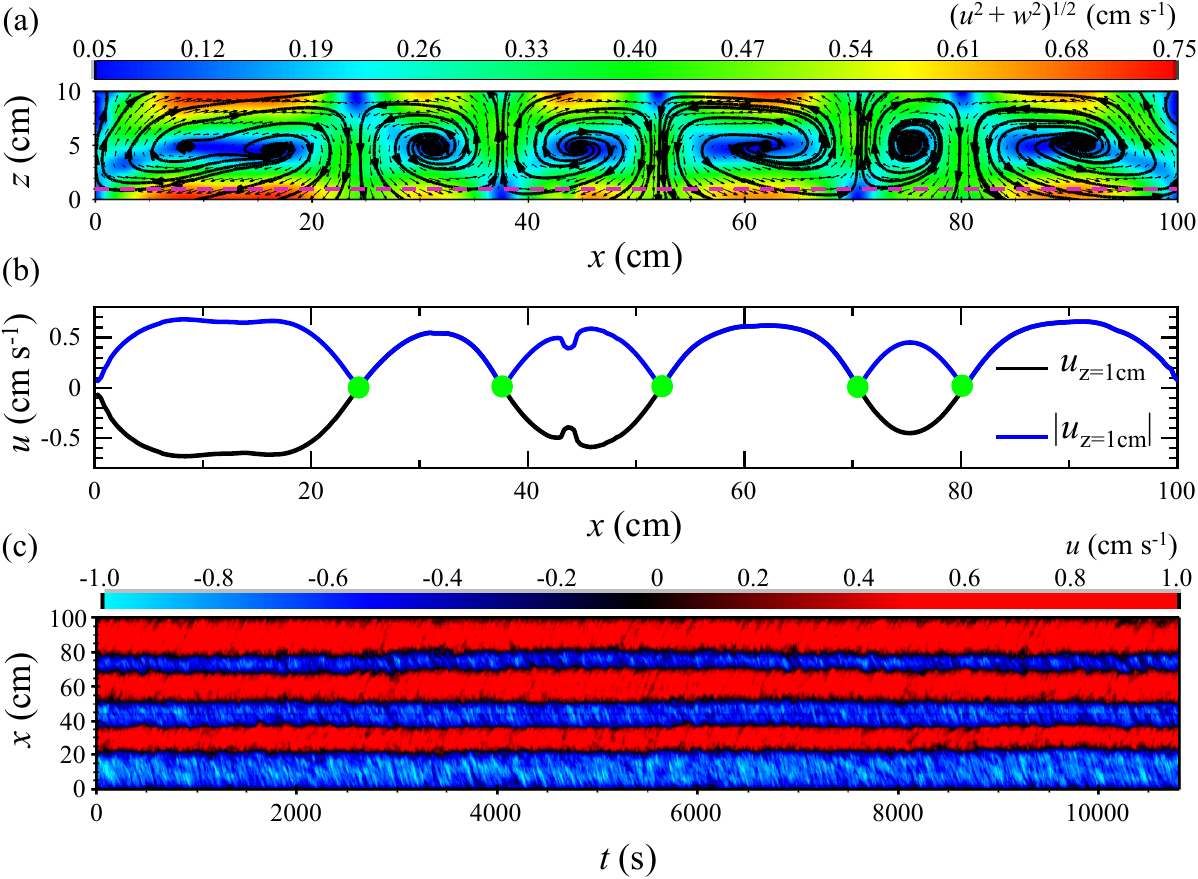}}
\caption{(a) Long time (three hours) averaged velocity field at $Ra = 3.5 \times 10^8, Pr = 9.9$. The color map represents the magnitude of the in-plane velocity $(u^2 + w^2)^{1/2}$.
(b) Horizontal velocity $u$ (black line) and its absolute value $\left\lvert u \right\rvert$ (blue line) along the line $z = 1 \text{cm}$.
Green dots mark the positions of local minimum in $\left\lvert u \right\rvert$, corresponding to roll boundaries.
(c) Space-time plot of horizontal velocity $u$ along $z = 1 \text{cm}$.}
\label{fig:f2}
\end{figure}

We first examine the characteristic flow structure in a cell with aspect ratio $\Gamma = 10$.
The flow typically organizes into multiple convection rolls aligned in the horizontal direction.
Among these configurations, a six roll structure is the most frequently observed one and emerges spontaneously within the present parameter space.
A representative long time averaged velocity field, obtained by averaging over three hours at $Ra = 3.5 \times 10^8$ and $Pr = 9.9$, is shown in \autoref{fig:f2}(a).
It is evident that the mean flow comprises six horizontally stacked convection rolls.
These rolls differ slightly in size as revealed by their horizontal extents and velocity distributions.
Small corner rolls appear near the upper left and upper right regions of the domain.
Such secondary corner rolls have also been reported in smaller aspect-ratio systems \citep{Xia2003PRE, Sugiyama2010PRL, Huang2016JFM, Chen2019JFM, Zhao2022JFM, Li2024JFM}.
A closer examination of the leftmost roll in \autoref{fig:f2}(a) shows the presence of internal substructures.
These features resemble the abnormal single roll state documented by \citet{Chen2019JFM, Chen2020JFM} and \citet{kar2020thermal}.
The maximum flow speed occurs near the top and bottom conducting plates.
This spatial velocity distribution is consistent with observations from flattened convection cells \citep{Qiu2001PRE, Xia2008conference, Li2022JFM, Li2024POF}.
Because the horizontal flow near the plates is confined to a narrower vertical section compared to the vertical flow near the sidewalls, conservation of mass requires both direction to transport equal volume flux.
As a result, the horizontal velocity near the conducting plates must be greater than the vertical velocity adjacent to the sidewalls.
The relation between roll size and characteristic velocity is examined in detail in \autoref{sec:Re}.
Another important feature is that adjacent rolls display horizontal velocities of opposite sign at the same vertical position.
This alternating pattern provides a convenient way to identify neighbouring rolls.
To quantify this spatial variation, we extract the horizontal velocity along a line located $1\ \text{cm}$ above the bottom plate, i.e., at $z = 1\ \text{cm}$.
This height is chosen because the horizontal velocity is nearly maximal at this location.
The resulting profile is shown in \autoref{fig:f2}(b), where both the horizontal velocity $u$ and its absolute value $\left\lvert u \right\rvert$ are plotted as functions of $x$.
Local minimum in $\left\lvert u \right\rvert$ correspond to roll boundaries and are marked by green dots, which provide a direct measure of the roll width.
The temporal evolution of the number, size, and position of the rolls is illustrated by the horizontal velocity map in \autoref{fig:f2}(c), where alternating color bands delineate the roll structures over time.

\begin{figure}
\centerline{\includegraphics[width=1\columnwidth]{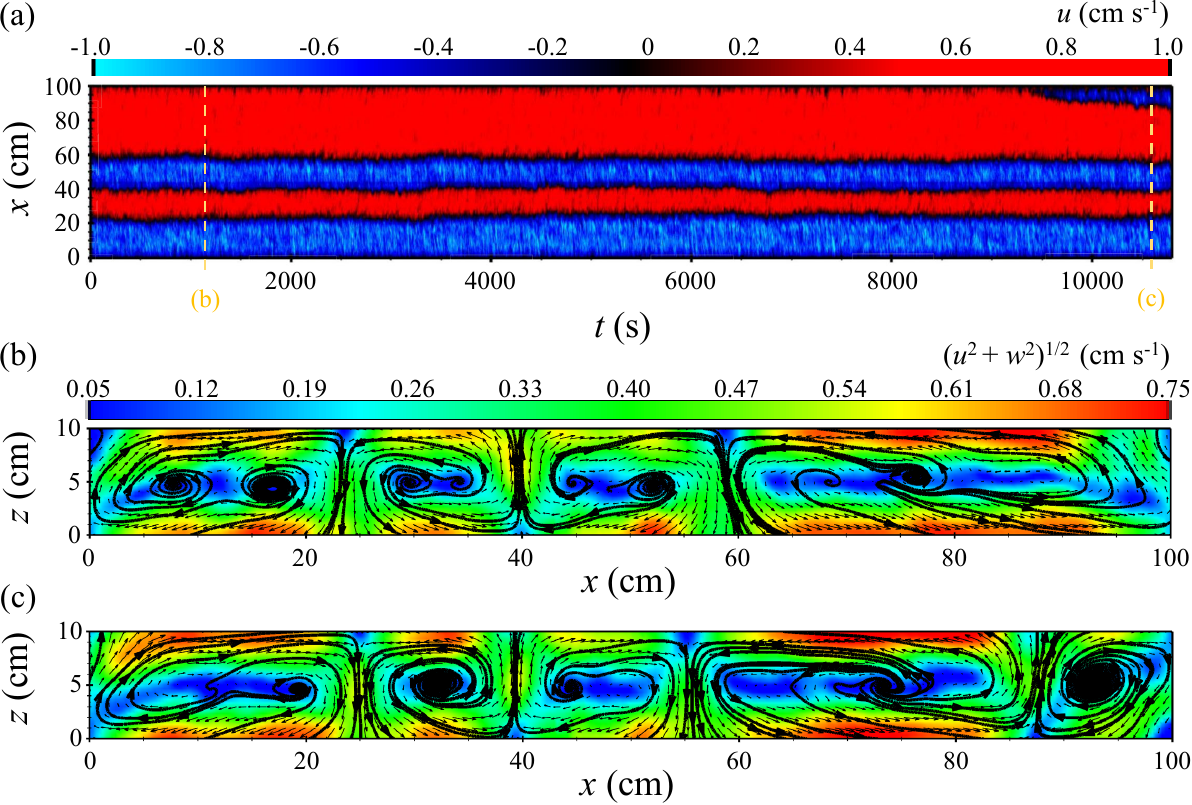}}
\caption{(a) Space-time plot of horizontal velocity $u$ along $z = 1 \text{cm}$ for $Ra = 3.5 \times 10^8, Pr = 9.9$.
Vertical dashed lines indicate the times at which short-time-averaged velocity fields are extracted.
(b, c) Velocity fields averaged over one turnover time $t_E$, showing (b) a four-roll state and (c) a five-roll state.
Color represents the in-plane velocity magnitude $(u^2 + w^2)^{1/2}$}
\label{fig:f3}
\end{figure}

In our experiments, the six-roll structure aligned horizontally is the most commonly observed and representative flow state in the $\Gamma = 10$ cell.
However, this configuration is not unique.
We find that for identical $(Ra, Pr)$ combinations the flow can spontaneously evolve into different states with varying numbers of horizontally stacked rolls.
It should be emphasized that for each realization, the working fluid was first kept in a quiescent state for at least six hours before initiating the experiment.
This procedure ensured that the system was free of residual motion and that the observed flow structures are not influenced by the flow history.
In \autoref{fig:f3}(a), we show the time series of horizontal velocity at $z = 1\ \text{cm}$ for $Ra = 3.5 \times 10^8, Pr = 9.9$.
From the figure one can see that the flow spontaneously evolves from four-roll state to five-roll state around $t= 9300\ \text{s}$.
Two representative snapshots of short time averaged velocity fields are displayed in \autoref{fig:f3}(b) and \autoref{fig:f3}(c), corresponding to four-roll and five-roll states, respectively.
The short time averaged velocity fields are obtained by averaging over one turnover time $t_E$.
The turnover time $t_E$ is estimated from the maximum instantaneous velocity and the average perimeter of an individual convection roll.
For an instantaneous flow field having a maximum horizontal velocity $u_0$, we define the average perimeter of a single roll as $C = ((n + 1) \times H + 2L) / n$, where $n$ is the number of convection rolls.
This yields an estimate $t_E = C / u_0$.
These stable configurations, namely the six rolls state shown in \autoref{fig:f2}(a), the four rolls state shown in \autoref{fig:f3}(b) and the five rolls state shown in \autoref{fig:f3}(c), clearly demonstrate the coexistence of multiple flow states under identical control parameters.
In each of the observed cases, the rolls are not in equal size, and the largest roll grows noticeably in size as the total number of rolls decreases.
The horizontal velocity near the plates remains significantly greater than the vertical velocity near the sidewalls.
These features closely resemble those reported in flattened convection cells \citep{Qiu2001PRE, Xia2008conference, Li2022JFM}.

\begin{figure}
\centerline{\includegraphics[width=1\columnwidth]{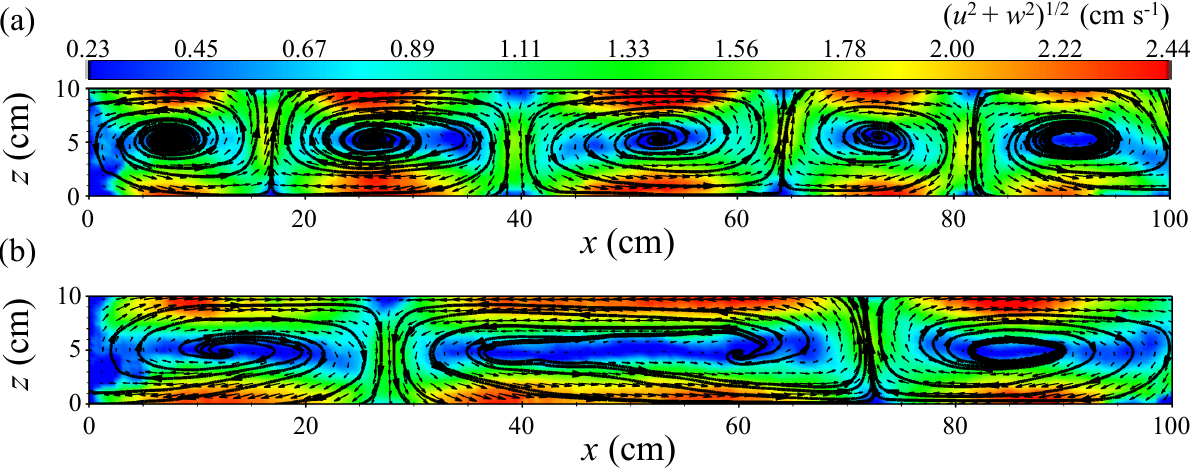}}
\caption{Short time averaged flow field at $Ra = 7.2 \times 10^9, Pr = 9.9$.
(a) Five-roll state. (b) Triple-roll state.}
\label{fig:f4}
\end{figure}

To further examine whether the occurrence of multiple flow states is related to the insufficient turbulence intensity, we increase the $Ra$ by more than one order of magnitude.
This is achieved by using 0.65 cSt silicone oil as the working fluid.
\autoref{fig:f4} shows representative short time averaged flow fields at $Ra = 7.2 \times 10^9$ and $Pr = 9.9$.
Even with a Rayleigh number $Ra$ that is about twenty times larger, the system still exhibits multiple stable flow states.
Specifically, a five-roll configuration is shown in \autoref{fig:f4}(a), and a triple-roll configuration is shown in \autoref{fig:f4}(b).
These observations confirm that multiple flow states persist even when the turbulence intensity is significantly increased.
The flow structures at this higher $Ra$ remain dominated by flattened vortices with different horizontal extent.
When comparing \autoref{fig:f2}, \autoref{fig:f3}, and \autoref{fig:f4}, a clear trend emerges: the number of rolls that can stably exist tends to decrease as the $Ra$ increases.
In our experiments with sufficiently high $Ra$, when the initial roll configuration is imposed through a controlled perturbation which will be explained in \autoref{sec:control}, the system can even support a single roll structure that spans the entire cell length.

\begin{figure}
\centerline{\includegraphics[width=1\columnwidth]{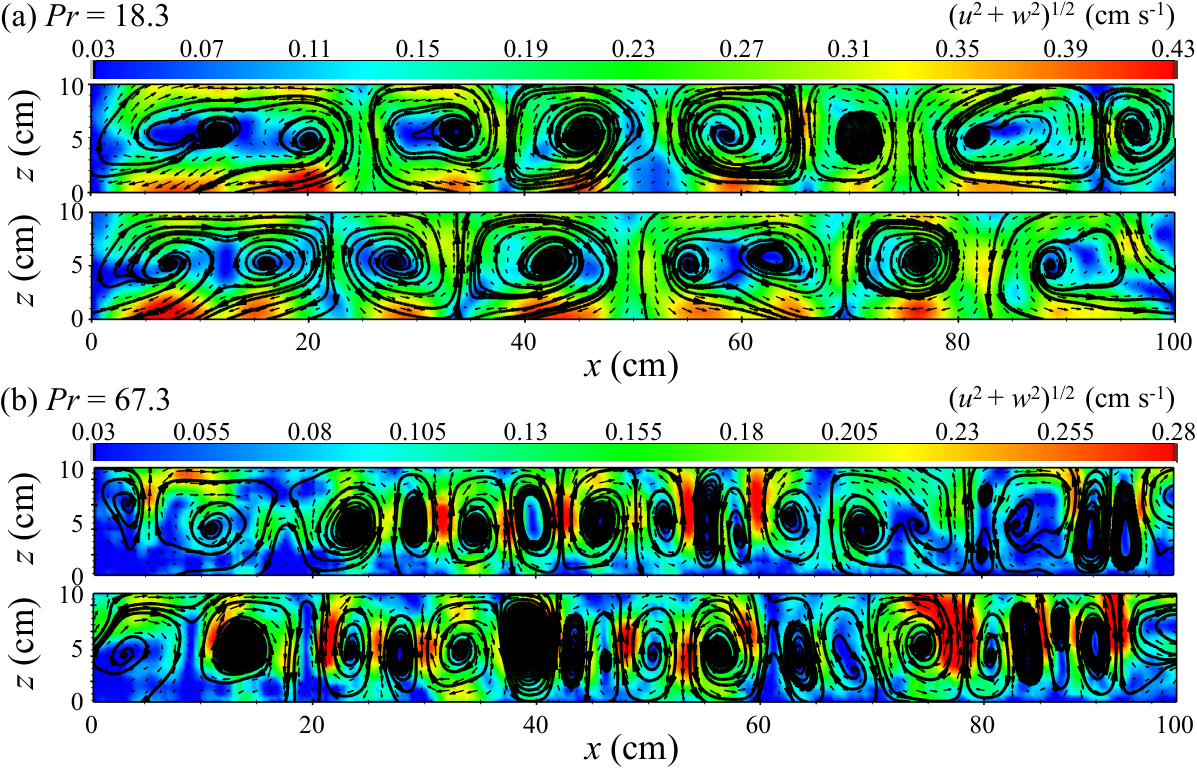}}
\caption{Multiple flow states at different $Pr$ numbers. (a) shows the seven-roll state and six-roll state at $Ra = 2.05 \times 10^8$, $Pr=18.3$, respectively. (b) shows the roll-like flow structures at $Ra = 1.63 \times 10^8$, $Pr=67.3$.}
\label{fig:f5}
\end{figure}

\begin{figure}
\centerline{\includegraphics[width=1\columnwidth]{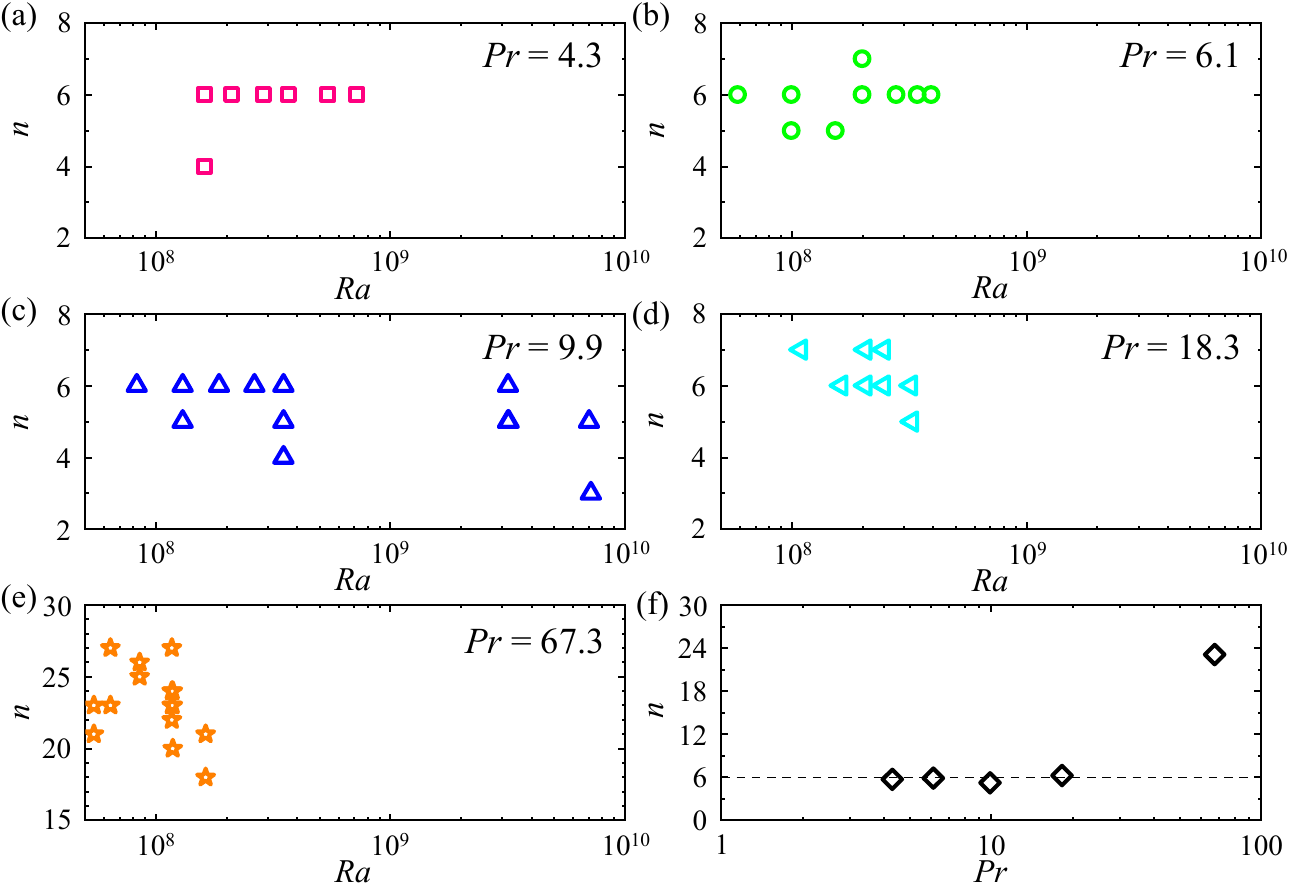}}
\caption{(a - e) Phase diagram showing the number of horizontally stacked rolls $n$ as a function of Rayleigh number $Ra$ for different Prandtl number $Pr$.
Each data point corresponds to an experimentally observed stable roll configuration.
(f) Arithmetic mean of the number of horizontally stacked rolls $n$ obtained from (a - e) as a function of the Prandtl number $Pr$.
The dashed line represents $n = 6$.
}
\label{fig:f6}
\end{figure}

To assess the generality of the multiple flow states, we also examine the flow field for different $Pr$.
Aqueous glycerol solutions of different concentrations are used to systematically adjust $Pr$.
We observe that multiple flow states occur over a wide range of $Pr$ from 4.3 to 67.3, which indicates that their coexistence is not confined to a specific value of $Pr$.
In \autoref{fig:f5}, we present typical flow fields for $Pr = 18.3$ and $Pr = 67.3$ at similar $Ra$.
For $Pr = 18.3$, the flow field exhibits a structure similar to that of the $Pr = 9.9$ cases discussed above.
When $Pr$ is further increased to 67.3, a clear difference is that the number of roll-like structures present in the flow field appears much larger than in the lower $Pr$ cases.
It should be noted that we use the term roll-like structure for the $Pr = 67.3$ case, since even with the aid of streamlines the flow structures are not as well defined as the rolls in the lower $Pr$ cases.
However, we would still count these roll-like structures as rolls when we analyse how the number of rolls changes with $Pr$.
Now, We can map out the phase diagram of the number of rolls in the parameter space of $Ra$ and $Pr$, as shown in \autoref{fig:f6}.
The phase diagrams in \autoref{fig:f6} (a) to (e) for different $Pr$ collect the number of rolls of the experimentally observed stable configurations and demonstrate the robust existence of multiple flow states in the $\Gamma = 10$ cell.
In \autoref{fig:f6} (f), we plot the arithmetic mean of the data points shown in the phase diagrams as a function of $Pr$.
The average number of rolls remains nearly unchanged for $4.3 \leq Pr \leq 18.3$ and increases significantly when $Pr$ reaches 67.3.
These results suggest a transition in the flow organization at large $Pr$.
There is a clear discrepancy between the lower $Pr$ cases and the case with $Pr = 67.3$ with respect to the flow field.
As stated in the previous paragraph, the largest velocity occurs near the horizontal plates and is primarily in the horizontal direction for the lower $Pr$ cases, whereas in \autoref{fig:f5}(b) for $Pr = 67.3$ the largest velocity is in the vertical direction between roll like flow structures.
This discrepancy implies that the flow may enter a new regime in which vertical motion dominates the flow field.
This behaviour is consistent with the flow field reported by \citet{wang2024prandtl} at high $Pr$, where the flow is dominated by thermal plumes rather than by the large-scale circulations.

Regarding the spontaneously occurring multiple states, our experiments reveal two scenarios.
One scenario is that the flow selects one flow state when it is initiated from a quiescent state.
The other scenario is that the flow spontaneously transitions from one well formed state, which remains stable for a long time, to another distinct state, as shown in \autoref{fig:f3}.
In our experiments, such spontaneous transitions from one state to another are rare, but they can still be observed occasionally.
We emphasize that the phase diagram in \autoref{fig:f6} does not represent an exhaustive enumeration of all possible flow states.
Instead, it includes only those stable configurations that have been identified in our experiments so far.
A comprehensive understanding of the full range of possible roll states will require further systematic investigation.
Such an investigation may be carried out most effectively with the aid of numerical simulations.

Although the presence of multiple flow states is consistent with findings from two-dimensional numerical simulations\citep{wang2020PRL,Wang2018PRF,Wang2020JFM}, the number of rolls that stably exist in experiments is significantly lower than that predicted in simulations. \citet{wang2020PRL} showed that imposing different initial conditions leads to different stable flow states in two-dimensional simulations and the mean aspect ratio of individual rolls, $\Gamma_r = \Gamma / n$, lies within $2/3 \leq \Gamma_r \leq 4/3$, where $n$ denotes the number of rolls.
According to this relation, a system with $\Gamma = 10$ should permit at least eight rolls, which exceeds the number observed in our experiments.
They also found that $\Gamma_r$ decreases as $Ra$ increases, implying that the number of rolls should grow with $Ra$.
In contrast, our experiments show that the number of rolls tends to decrease with $Ra$.
For example, for $Pr = 9.9$, one can see that with increasing $Ra$ the number of rolls can decrease to three.
This implies that the discrepancy between the numerical results and the experimental observations is expected to become more pronounced at higher $Ra$.
The simulations employed horizontal periodic boundary conditions which implies infinite long in horizontal direction while our experiments have rigid walls.
It is therefore reasonable to conjecture that the difference in the number of rolls may arise from the different boundary conditions and the sidewall boundary condition appears to play an essential role in determining the number of stable rolls.
This interpretation is supported by \citet{Wang2018PRF}, who found that the number of stable rolls decreases when the periodic boundary condition is replaced by solid sidewalls.
Specifically, for $Ra = 10^7$ and $Pr = 0.71$, they reported 5–8 rolls for $\Gamma = 8$ and 7–12 rolls for $\Gamma = 12$.
Even for the same boundary conditions, the number of stable rolls in simulations remains higher than in experiments.
A more plausible explanation lies in the difference in dimensionality between simulations and experiments.
Experimental systems are inherently three-dimensional, and the lateral extent of the cell, denoted by $\Gamma_{\text{lateral}} = W / H$, introduces additional disturbance that are absent in 2D simulations.
For comparable Rayleigh numbers ($Ra \sim 10^8$), the influence of $\Gamma_{\text{lateral}}$ can be assessed from previous studies.
In the current experiments with $\Gamma_{\text{lateral}} = 0.3$, the number of spontaneously generated rolls were predominantly in the range of 4 to 7, corresponding to the $\Gamma_r$ in a range of $10/7 \leq \Gamma_r \leq 5/2$.
\citet{Xia2008conference} reported that in a rectangular cell with $\Gamma = 9.9$ and $\Gamma_{\text{lateral}} = 2.5$, the turbulent structure consisted of three horizontally stacked rolls, with the average size of the roll having $\Gamma_r \approx 3.3$.
In systems with even larger $\Gamma_{\text{lateral}}$, often referred to as mesoscale convection, the flow develops into turbulent superstructures with substantially larger roll sizes, typically $\Gamma_r \geq 6$ \citep{Pandey2018NC, Stevens2018PRF, Green2020JFM, Krug2020JFM, Pandey2022JFM, Moller2022JFM}.
These observations consistently indicate that roll size increases with the expansion of the third dimension.
This provides a straightforward explanation for the observed discrepancy between experiments and two-dimensional simulations regarding both the number and the size of roll structures.

\subsection{Reynolds number}\label{sec:Re}
An interesting question concerning the multiple-roll configuration in a $\Gamma = 10$ cell is whether a systematic relationship exists between $Re$ and the size of the rolls.
To investigate this relationship, we first extract the individual rolls using the method illustrated in \autoref{fig:f2}(b).
The size of each roll is then determined, and all velocity data enclosed within its boundaries are extracted.
Once the roll sizes are obtained, the aspect ratio of each roll is defined as $\Gamma_{\text{roll}} = W_{\text{roll}}/H$, where $W_{\text{roll}}$ denotes the roll width and $H$ is the cell height.
All rolls in our experiments share the same height $H$.
Only cases in which the rolls remain stable for at least one hour during data acquisition are included in the analysis.
The stability of the rolls is assessed from the space-time plot of the horizontal velocity $u$ along $z = 1\ \text{cm}$, for example \autoref{fig:f2}(c), which shows no significant variation for at least one hour.

\begin{figure}
\centerline{\includegraphics[width=1\columnwidth]{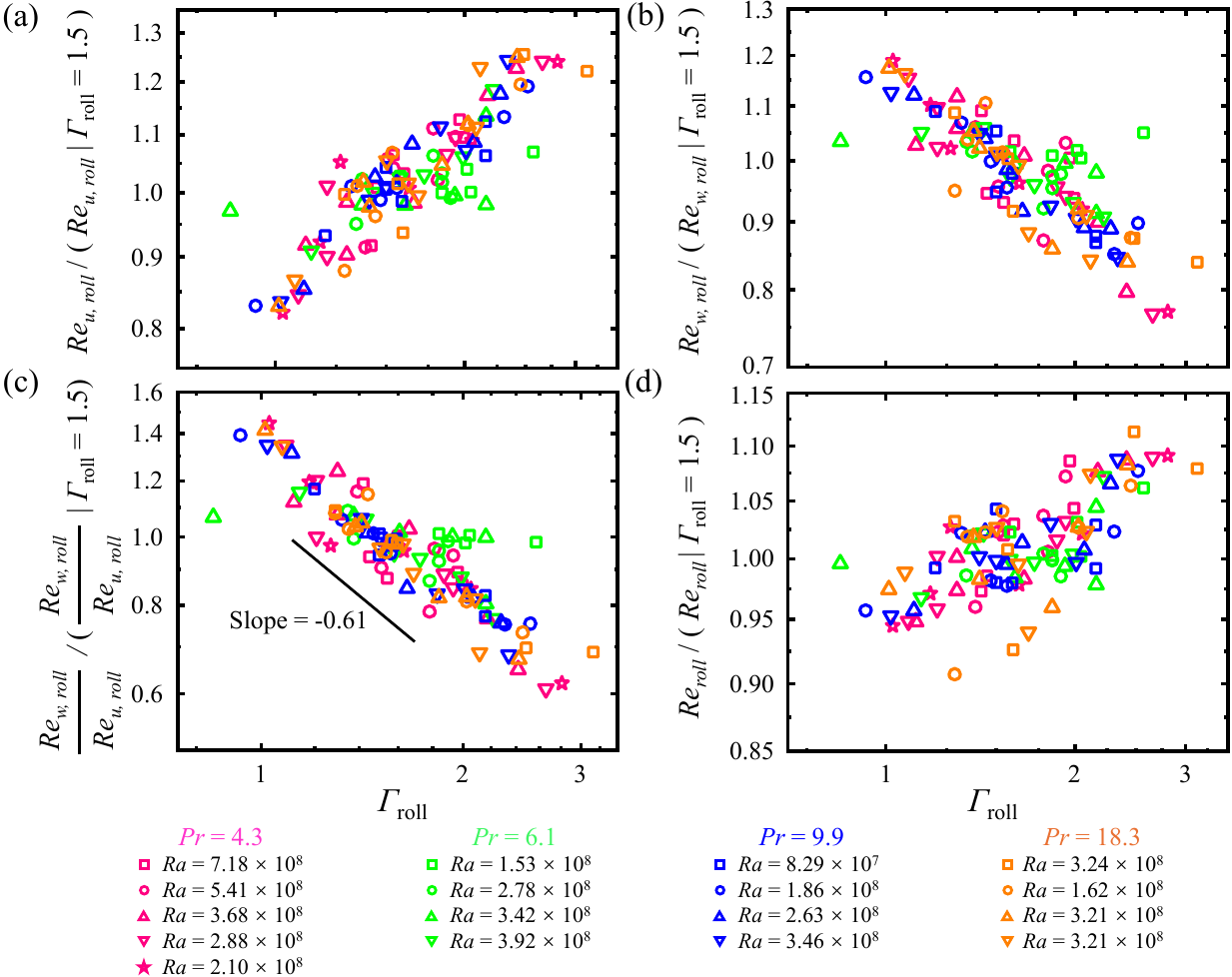}}
\caption{$Re$ as function of $\Gamma_{\text{roll}}$ at different $Ra$ and $Pr$.
(a) $Re_{u,\text{roll}} = \sqrt{\langle u^2 \rangle_{\text{roll}}} H / \nu$ based on horizontal velocity.
(b) $Re_{w,\text{roll}} = \sqrt{\langle w^2 \rangle_{\text{roll}}} H / \nu$ based on vertical velocity.
(c) $Re_{w,\text{roll}} / Re_{u,\text{roll}}$ (component ratio).
(d) $Re_{\text{roll}} = \sqrt{\langle u^2 + w^2 \rangle_{\text{roll}}} H / \nu$ based on total velocity.
All data are normalized by their corresponding values at $\Gamma_{\text{roll}} = 1.5$.}
\label{fig:f7}
\end{figure}

\autoref{fig:f7} presents the dependence of $Re$ on $\Gamma_{\text{roll}}$.
Although the data are scattered across the various experimental conditions, the overall trends are consistent.
The scatter likely reflects the limited number of datasets.
Panels (a), (b), and (d) display the Reynolds numbers based on horizontal velocity $Re_{u,\text{roll}} = \sqrt{\langle u^2 \rangle_{\text{roll}}} H / \nu$, vertical velocity $Re_{w,\text{roll}} = \sqrt{\langle w^2 \rangle_{\text{roll}}} H / \nu$, and total in-plane velocity $Re_{\text{roll}} = \sqrt{\langle u^2 + w^2 \rangle_{\text{roll}}} H / \nu$, respectively.
$\langle \rangle_{\text{roll}}$ denotes the average taken within the boundary of each individual roll.
Panel (c) shows the ratio $Re_{w,\text{roll}} / Re_{u,\text{roll}}$ as a function of $\Gamma_{\text{roll}}$.
To collapse the data obtained under different $Ra$ and $Pr$, all Reynolds numbers are normalized by their corresponding values at $\Gamma_{\text{roll}} = 1.5$.
The values at $\Gamma_{\text{roll}} = 1.5$ are not always available in the raw records.
These values are therefore obtained by linear interpolation between the contiguous datasets that bracket $\Gamma_{\text{roll}} = 1.5$.
As shown in \autoref{fig:f7}(a), $Re_{u,\text{roll}}$ increases with increasing $\Gamma_{\text{roll}}$.
In contrast, $Re_{w,\text{roll}}$ decreases monotonically with roll size, as shown in \autoref{fig:f7}(b).
This opposite behavior can be understood from mass conservation.
Assuming incompressibility, the horizontal and vertical volume fluxes must balance, implying $u H \sim w l$, and hence $w/u \sim 1/\Gamma_{\text{roll}}$.
Accordingly, larger rolls require higher horizontal velocities and lower vertical velocities to maintain mass balance.
% From a physical standpoint, two adjacent rolls of different sizes share a common vertical velocity at their interface.
% Thus, the larger roll must possess a stronger horizontal flow to conserve mass, leading to a larger horizontal $Re$.
% In contrast, the vertical velocity is distributed over a larger area in the larger roll, resulting in a smaller vertical $Re$.
This reciprocal relationship is captured in \autoref{fig:f7}(c), where the ratio $Re_{w,\text{roll}} / Re_{u,\text{roll}}$ decreases systematically with increasing $\Gamma_{\text{roll}}$.
The observed trend is consistent with two-dimensional numerical simulations of \citet{wang2020PRL}.
A linear fit to the data in \autoref{fig:f7}(c) yields a scaling exponent of $-0.61 \pm 0.03$, which is very close to the exponent of $-0.68$ reported in simulations of \citep{Wang2020JFM}.
The deviation of current experiments from the theoretical prediction of $w/u \sim 1/\Gamma_{\text{roll}}$ can be attributed to the three-dimensional nature of the flow, the presence of secondary corner rolls, and the mass exchange across adjacent rolls.
As shown in \autoref{fig:f7}(d), the total Reynolds number $Re_{\text{roll}}$ tends to increase slightly with roll size.
This further indicates that the quasi-two-dimensional flow in the $\Gamma = 10$ cell is primarily driven by horizontal motion as the the total Reynolds number $Re_{\text{roll}}$ shares the similar trend with $Re_{u,\text{roll}}$.
Due to the absence of full-field temperature measurements, we are unable to determine the relationship between the $Nu$ and roll size.

\begin{figure}
\centerline{\includegraphics[width=1\columnwidth]{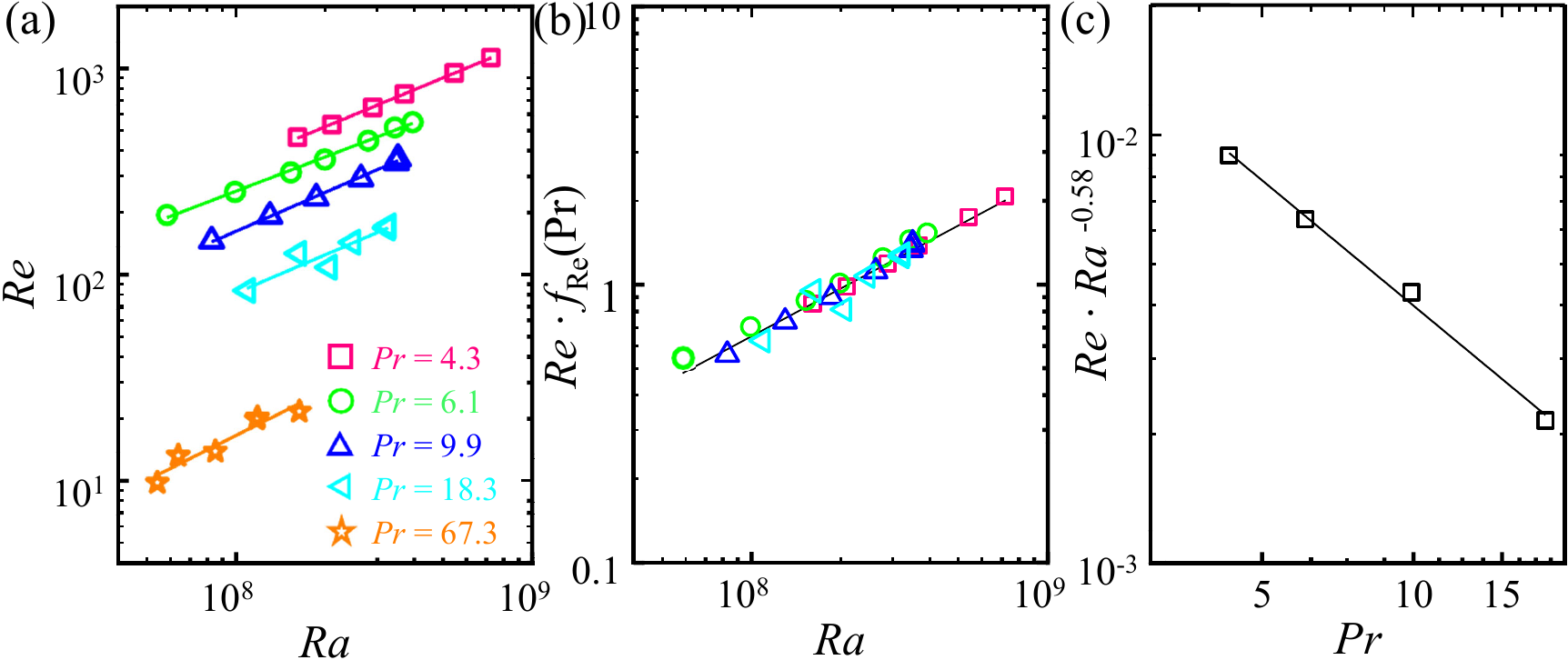}}
\caption{(a) $Ra$ dependence of $Re$ for different $Pr$.
(b) Normalized $Re$ data, scaled by the value at $Ra = 2 \times 10^8$.
(c) $Pr$ dependence of the compensated $Re$.}
\label{fig:f8}
\end{figure}

We further investigate the dependence of global Reynolds number $Re$ on $Ra$ across different $Pr$ using the full-field velocity data.
The global Reynolds number is defined as $Re = \langle \sqrt{u^2 + w^2} \rangle H / \nu$, where $\langle \rangle$ denotes the average over the entire flow field.
\autoref{fig:f8} shows that $Re$ decreases monotonically with increasing $Pr$, and this decrease becomes particularly pronounced as $Pr$ increases from 18.3 to 67.3.
This trend coincides with a significant structural transition in the flow field.
At $Pr = 67.3$, the roll structure is no longer dominant.
Instead, the flow is characterized by vertically aligned plumes.
%This marks a qualitative change from the roll-dominated circulation observed at lower $Pr$ values.
In addition to the structural transition, there is a clear change in the scaling behavior of the Reynolds number.
As indicated in \autoref{table_ReRa}, the scaling exponent $\gamma$ in the relation $Re \sim Ra^\gamma$ remains nearly constant for $4.3 \leq Pr \leq 18.3$, and agrees well with previous experiments and simulations conducted in quasi-2D or 2D systems with $\Gamma = 1$ \citep{Sugiyama2009JFM, vanderPoel2013JFM, Zhang2017JFM, wang2020PRL, Li2021JFM}.
When $Pr$ increases from 18.3 to 67.3, the exponent $\gamma$ increases significantly.
A similar transition has been reported in both three-dimensional \citep{Lam2002PRE, Silano2010JFM} and quasi-2D \citep{Li2021JFM} cell with $\Gamma = 1$.
In particular, the experiments of \citet{Li2021JFM} revealed that a sharp increase in the scaling exponent corresponds to the breakdown of large-scale circulation.
This is consistent with the transition observed in the present study.
It is worth noting that in \citet{Li2021JFM}, the structural transition occurred at a much higher $Pr \approx 345$, which differs from our result.
% This discrepancy likely stems from the difference in $Ra$.
% Their experiments were conducted at significantly higher $Ra$, which implies stronger buoyancy forcing.
% At higher $Ra$, the buoyancy force is capable of overcoming larger viscous resistance.
% As a result, the transition to plume-dominated small-scale structures is postponed to higher $Pr$.
% In contrast, at lower $Ra$, the viscous damping becomes more effective, and the flow transitions earlier to a regime dominated by small-scale plume dynamics.
\citet{Chong2016JFM} investigated the flow field in geometrically confined domains and found that reducing the lateral aspect ratio $\Gamma_{\text{lateral}}$ leads to distinct changes in the dominant flow regimes.
As $\Gamma_{\text{lateral}}$ decreases, the flow transitions sequentially from the classical boundary-layer-controlled regime, to a plume-controlled regime, and eventually to a severely confined regime.
These changes are accompanied by a progressive increase in the scaling exponent of the $Re$.
Upon entering the third regime, the large-scale circulation, which is typically formed by mushroom-like and fragmented thermal plumes, breaks down.
The flow becomes dominated by long lived, vertically aligned finger like plume columns, and the scaling exponent of the $Re$ associated with this plume dominated state matches the value observed in the present experiment.
This agreement provides further evidence that sudden changes in either the magnitude or the scaling exponent of the Reynolds number serve as reliable indicators of underlying transitions in the flow structure.

\begin{table}
\begin{center}
\def~{\hphantom{0}}
\begin{tabular}{ccccccccc}
$Pr$ &$ C_1$ & $\gamma_1$  \\[3pt]
4.3  & $5.72\times 10^{-3}$ & 0.60$\pm$0.01 \\
6.1  & $9.77\times 10^{-3}$ & 0.55$\pm$0.02 \\
9.9  & $1.55\times 10^{-3}$ & 0.63$\pm$0.02 \\
18.3 & $9.62\times 10^{-4}$ & 0.62$\pm$0.09 \\
67.3 & $2.65\times 10^{-5}$ & 0.72$\pm$0.10 \\
\end{tabular}
\caption{Fitted coefficients $C_1$ and scaling exponents $\gamma_1$ for the power-law relation $Re = C_1 Ra^{\gamma_1}$ at different $Pr$.}
\label{table_ReRa}
\end{center}
\end{table}

We then extract the scaling relationship of the $Re$ with both $Ra$ and $Pr$ by a commonly used method \citep{Xia2002PRL, Xie2017JFM, Li2021JFM}.
Specifically, to investigate the combined scaling $Re \sim Ra^\gamma Pr^\beta$, we first normalize the $Re$ values in each $Pr$ group using a constant that depends on $Pr$.
Here, the normalization factor is taken to be the $Re$ value at $Ra = 2 \times 10^8$ for each $Pr$.
This procedure collapses the data across different $Pr$ values, as shown in \autoref{fig:f8}(b).
By fitting the normalized data, we extract the scaling exponent $\gamma$ in the relation $Re \sim Ra^\gamma$.
Next, we compensate the original $Re$ values by dividing them by $Ra^\gamma$ and compute the average within each $Pr$ group.
This allows us to determine the scaling exponent $\beta$ in the relation $Re \sim Pr^\beta$.
It is important to note that data for $Pr = 67.3$ are excluded from this analysis due to significant changes in the underlying flow structure.
From the compensated data in \autoref{fig:f8}(b), we find that $Re$ follows the scaling $Re \sim Ra^{0.58}$.
%This exponent is considerably larger than typical values reported in three-dimensional cell, where $\gamma$ generally lies in the range $0.4$ to $0.5$.
The enhanced scaling observed here is consistent with stronger momentum transport in two-dimensional or quasi-two-dimensional cells, as previously noted by \citet{vanderPoel2013JFM}, \citet{Zhang2017JFM} and \citet{Chen2020JFM}.
The $Pr$ dependence of $Re$ is shown in \autoref{fig:f8}(c), where the data have been compensated by $Ra^{0.58}$.
The resulting scaling exponent is $\beta = -0.97 \pm 0.01$.
This value agrees well with previous experimental and numerical results for quasi-two-dimensional cells with $\Gamma = 1$ \citep{vanderPoel2013JFM, Li2021JFM, he2021CPB}.
It is worth noting that prior studies of small aspect ratio systems ($\Gamma \leq 1$) have reported a wide range of exponents for the $Re \sim Pr^\beta$ relationship, typically spanning from $-0.72$ to $-1$ \citep{Lam2002PRE, verzicco1999JFM, Breuer2004PRE, Brown2007JSM, Yang2020POF}.
Our current findings fall within this range.

\subsection{The flow control and the heat transfer}\label{sec:control}

\begin{figure}
\centerline{\includegraphics[width=1\columnwidth]{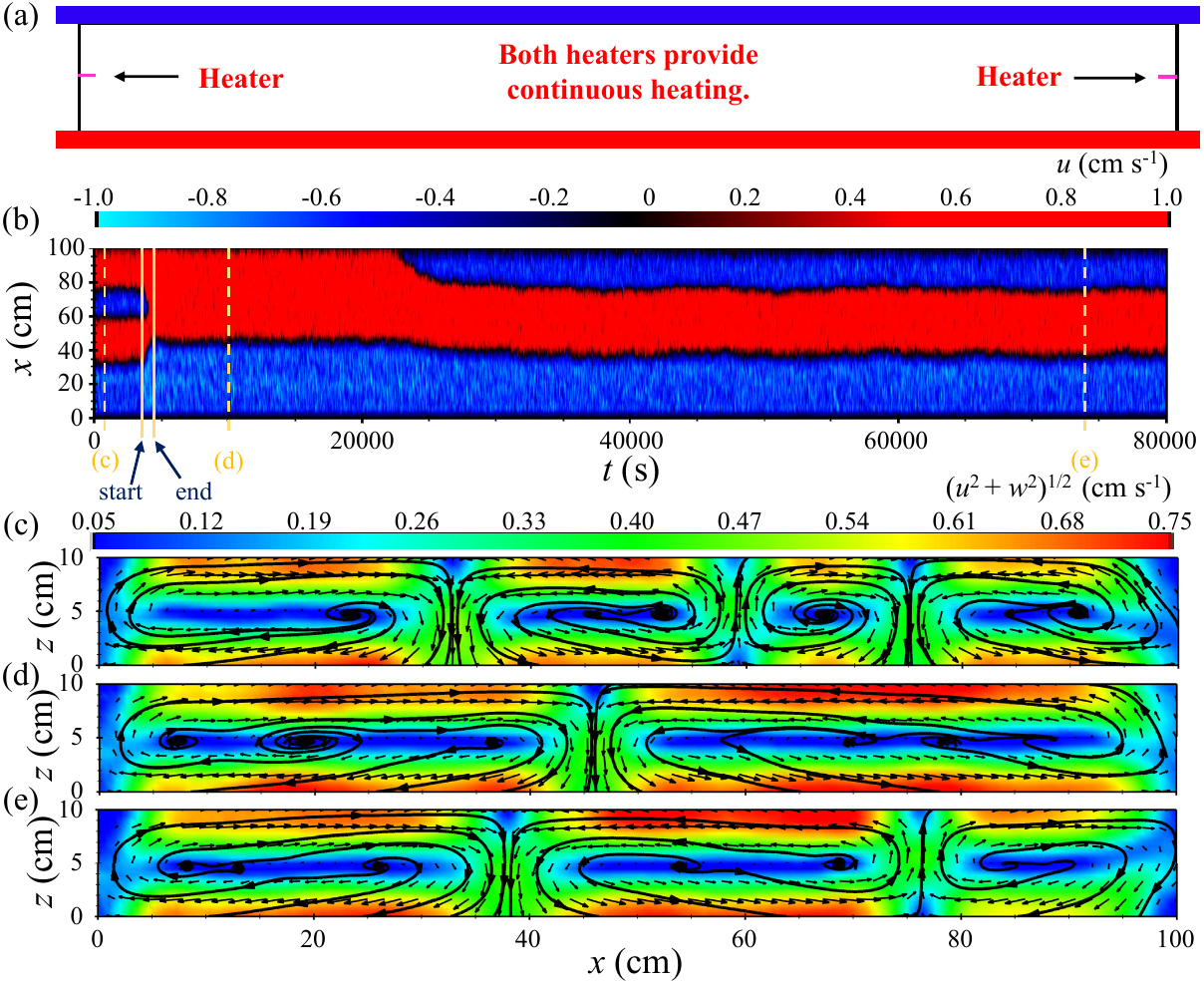}}
\caption{(a) Schematic showing initial flow field generated by local sidewall heating.
The heaters are positioned 5 mm away from the centre plane and therefore do not interfere with the PIV measurement.
Each heater has a resistance of $5.6~\Omega$, a diameter of $4~\text{mm}$, and extends $10~\text{mm}$ into the fluid domain.
(b) Time evolution of horizontal velocity at $z = 1~\text{cm}$ for $Ra = 3.5 \times 10^8$, $Pr = 9.9$.
The two solid lines denote the start and end of the heating period.
(c – e) Representative flow structures observed during different stages of long time evolution.}
\label{fig:f9}
\end{figure}

Inspired by numerical simulations, in which different initial conditions can be imposed on the flow field, we next examine how the flow responds to controlled perturbations of its initial state.
Here the controlled perturbation is achieved by localized heating that uses two electric heaters placed at the mid height of the left and right sidewalls, as can be seen in \autoref{fig:f9}(a).
\autoref{fig:f9}(b) shows the time evolution of the horizontal velocity at $z = 1~\text{cm}$ for $Ra = 3.5 \times 10^8$ and $Pr = 9.9$, spanning a total duration of approximately 22.2 hours.
In the present experiments, the flow field self-organises into a four roll state when started from a quiescent condition, as shown in \autoref{fig:f9}(c).
When the heaters are activated, as indicated by the first solid line, the two outermost rolls begin to expand and compress the adjacent inner rolls.
As heating continues, these rolls merge to form two large flattened rolls of approximately $L/2$ in length, which leads to a dominant double roll structure shown in \autoref{fig:f9}(d), where the left and right rolls rotate clockwise and counterclockwise.
This double roll state remains stable for about 5.1 hours after the heaters are switched off.
This double roll state has not been observed in the previous experiments.
Subsequently, a corner roll begins to grow on one side of the domain and the system evolves into a stable triple roll state, as shown in \autoref{fig:f9}(e).
The resulting triple roll state is also a previously unobserved flow state.
As we emphasize in the previous section, the phase diagram shown in \autoref{fig:f6} does not represent all possible flow states within the present range of $Ra$ and $Pr$, and the newly observed double and triple roll states clearly support this statement.
When discussing multiple flow states, one usually adopts the picture of potential well.
The phase diagram shown in \autoref{fig:f6} displays several states that are reached easily, which can be interpreted as potential wells with relatively large basins of attraction in the space of flow configurations.
The double and triple roll states identified here suggest the presence of additional, more remote potential wells that are much more difficult to access under natural fluctuations.
Furthermore, as mentioned above, spontaneous transitions from one state to another are quite rare in our experiments, which may imply that once the flow resides in one potential well, a considerable perturbation is required for it to escape to another well.
This interpretation is consistent with the level of energy input that is needed to alter the four roll state into a double roll state.
To successfully establish the double roll state, a heating power of $36.5~\text{W}$ is continuously applied for ten minutes, as shown in \autoref{fig:f9}(b).
Within this framework, the depth of a potential well reflects the robustness of the corresponding flow state to external disturbances and internal fluctuations.
The long residence times of the double and triple roll states indicate that these states correspond to comparatively deep wells.
The fact that the double roll and triple roll states are only reached after strong and localized perturbations suggests that their wells are narrow in phase space, so that the probability of entering them through spontaneous fluctuations is very small.
In contrast, the states that appear frequently in the phase diagram of \autoref{fig:f6} are likely associated with wider basins of attraction, which can be reached from many different realizations of the initial condition and noise history.
A more quantitative characterization of this potential well picture would require systematic control of the initial conditions and perturbations, which could be realized in future experiments and in direct numerical simulations.
Such studies would help to map out the relative depths and widths of the wells that correspond to different roll states and would clarify the mechanisms that govern transitions between them.

\begin{figure}
\centerline{\includegraphics[width=1\columnwidth]{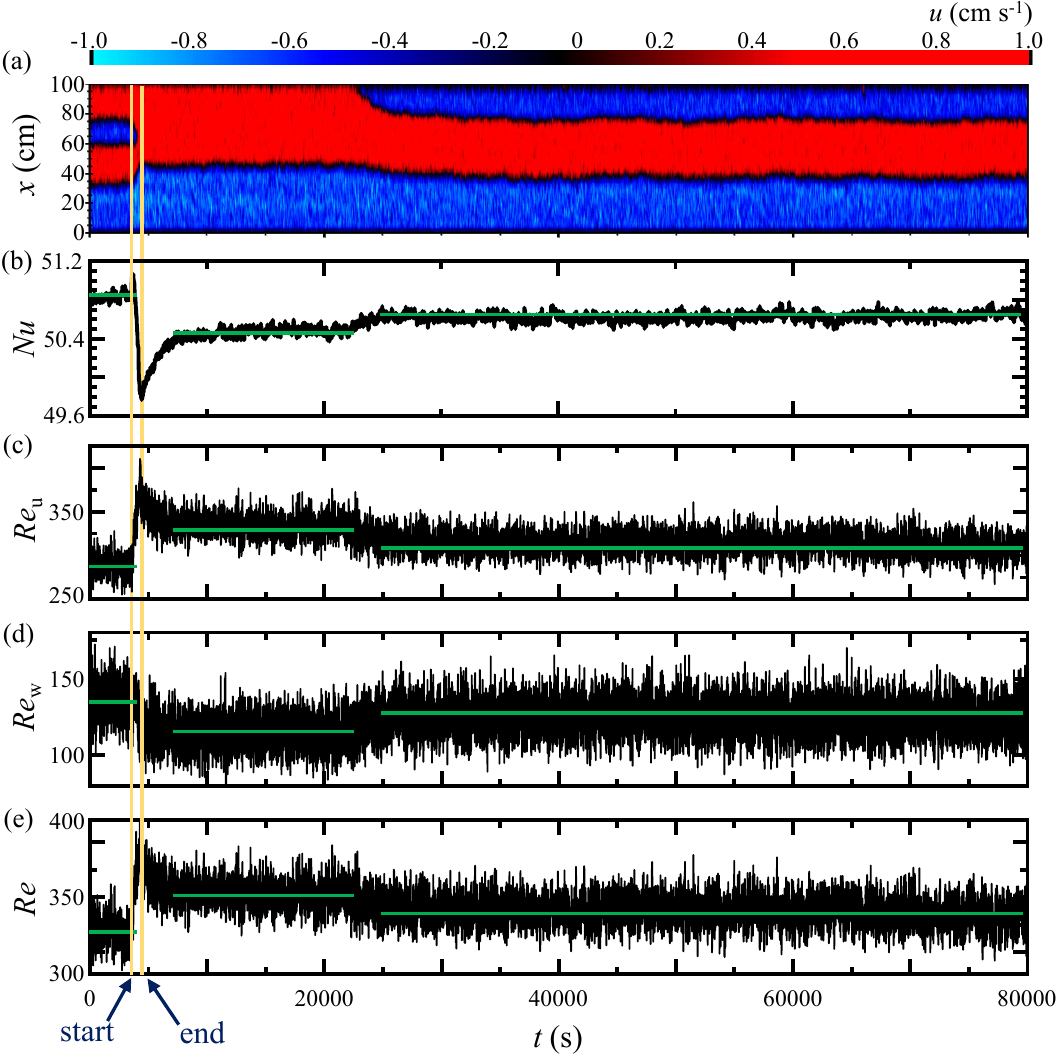}}
\caption{(a) Time evolution of flow structure.
(b) Time evolution of the Nusselt number $Nu$.
(c) Time evolution of the Reynolds number based on horizontal velocity $Re_u$.
(d) Time evolution of the Reynolds number based on vertical velocity $Re_w$.
(e) Time evolution of the global Reynolds number $Re$.
Here the $Re_u$, $Re_w$ and $Re$ are calculated over the full flow field.
Vertical lines mark the heater activation and deactivation times.
Horizontal lines in panels (b) to (e) indicate the different values associated with different flow states.}
\label{fig:f10}
\end{figure}

\autoref{fig:f10} presents the time evolution of the global heat and momentum transfer during this process, based on the same dataset used in \autoref{fig:f9}.
The vertical solid lines denote the start and the end of the heating period respectively.
Here the global heat transfer is quantified by the parameter Nusselt number $Nu$, which is calculated as $Nu = J / (\chi \Delta T /H)$, where $J$ and $\chi$ are the total heat flux entering the convection cell and the thermal conductivity of working fluid.
During the experiments, the total heat flux $J$ is kept constant, which is ensured by the long term stable power supply, and the small variations in $Nu$ shown below are due to changes in $\Delta T$ that reflect changes in the global heat transfer efficiency.
Initially, the system contains four rolls and exhibits a global Nusselt number of $Nu = 50.8$.
Upon transitioning to the double-roll state, the heat transfer decreases to a value of $Nu = 50.48$.
As the system evolves into the final triple-roll state, the Nusselt number rises slightly, from $Nu = 50.48$ to $Nu = 50.63$.
This increase indicates that the heat transfer grows with the number of horizontally stacked rolls.
This trend is consistent with the expectation that additional rolls provide more channels for thermal transport between the hot and cold plates.
One may notice that within the interval marked by the two vertical lines, the $Nu$ first increases then undergoes a sharp decline.
The initial increase results from thermal plumes generated by the heaters directly impinging on the top plate and propagating along it.
Although the top plate temperature is regulated by a refrigerated circulator, the system cannot remove the large, suddenly injected heat flux rapidly enough.
As a consequence, the top-plate temperature increases, reducing the imposed temperature difference $\Delta T$ and thereby producing a rise in $Nu$.
The subsequent drop in $Nu$ also stems from the additional heat delivered by the heaters.
Because the cooling capacity at the top plate is limited, the excess heat eventually raises the bottom-plate temperature, increasing $\Delta T$.
This leads to a pronounced decrease in $Nu$ during the heating phase.
Once the heating is terminated and the excess heat dissipates, the Nusselt number gradually recovers.
In parallel, the Reynolds number based on horizontal velocity $Re_u$ and the global Reynolds number $Re$ show a decrease with increasing number of rolls, while Reynolds number based on vertical velocity $Re_w$ exhibits an increasing trend, as shown in \autoref{fig:f10}(c–e).
Here the $Re_u$, $Re_w$ and $Re$ are all calculated over the full flow field.
These results agree with the analysis in \autoref{sec:Re}, where it was found that $Re_{u,\text{roll}}$ decreases and $Re_{w,\text{roll}}$ increases as roll size decreases.
Although we cannot directly assess the dependence of heat transfer on roll size $\Gamma_r$ due to the lack of spatially resolved temperature data, the consistency between the behavior of $Nu$ and $Re_w$ suggests a link between vertical momentum transport and thermal flux.
Comparing the trends of heat and momentum transfer, we find that $Nu$ is negatively correlated with $Re_u$ and positively correlated with $Re_w$.
This behaviour agrees with the results of previous two dimensional numerical simulations \citep{vanderPoel2012POF, Wang2020JFM}.

\section{Conclusions}\label{sec:conclusions}
In this study, we have conducted detailed experiments to investigate the turbulent flow structure in turbulent thermal convection with a large aspect ratio of $\Gamma = 10$.
The Rayleigh number $Ra$ spans from $5.4 \times 10^7$ to $7.2 \times 10^9$, and the Prandtl number $Pr$ ranges from $4.3$ to $67.3$.
The flow field is found to consist of multiple convection rolls aligned in the horizontal direction.
Repeated experimental runs under identical control parameters reveal the existence of multiple flow states, which confirms that the system possesses several distinct and stable flow configurations.
By imposing the localized sidewall heating, we find another two flow states, the double and triple roll state, which indicates that the final turbulent state depends sensitively on the perturbation or the initial condition.
The experiments also reveal rare but clear spontaneous transitions between different roll states, analogous to those reported previously in systems with small aspect ratio $\Gamma \leq 1$, which supports the view that multistability in turbulent thermal convection is a robust and generic feature.
The dependence of momentum transport on roll size $\Gamma_{\mathrm{roll}}$ has been quantified.
The Reynolds number based on the horizontal velocity within individual rolls, $Re_{u,\mathrm{roll}}$, increases with $\Gamma_{\mathrm{roll}}$, whereas the Reynolds number based on the vertical velocity, $Re_{w,\mathrm{roll}}$, and the ratio $Re_{w,\mathrm{roll}}/Re_{u,\mathrm{roll}}$ decrease with $\Gamma_{\mathrm{roll}}$.
These opposite trends can be explained by mass conservation between the horizontal and vertical flows within each roll.
The global Reynolds number $Re$ scales as $Re \sim Ra^{0.58}Pr^{-0.97}$ for $Pr \leq 18.3$, and a distinct change in magnitude and scaling exponent occurs when $Pr$ is increased to $67.3$, which reflects a structural transition from a roll dominated to a plume dominated flow organization.
We have also explored the relationship between the number of rolls and the global heat and momentum transfer.
An increase in the number of horizontally stacked rolls enhances the number of vertical transport channels between the hot and cold plates and thereby increases the overall heat flux.
At the same time, the vertical component of the momentum transfer strengthens, while the horizontal component weakens as the roll number increases.
These results show that both the global transport properties and the associated scaling behaviours are closely linked to the organization of the roll structures.
The observed states and transitions can be interpreted within a potential well picture in the space of flow configurations.
States that appear frequently in the phase diagram correspond to wells with wide basins of attraction, which are easily reached from many realizations of the initial condition and noise history.
In contrast, the double and triple roll states generated by strong localized heating are associated with more remote wells that are narrow in phase space, which explains why they are not accessed by natural fluctuations and why spontaneous transitions between states are rare.
Overall, our findings provide a comprehensive experimental view of the complex flow organization and of the associated transport mechanisms in turbulent thermal convection with large aspect ratio.
The results demonstrate that the emergence of multiple flow states is an intrinsic feature of such systems and that transitions among these states can significantly influence global heat and momentum transport.
Future investigations combining fully resolved three-dimensional measurements and high fidelity simulations will be crucial for clarifying the underlying mechanisms governing state selection, stability, and transitions in large-scale turbulent thermal convection.

\backsection[Funding]
{This work was supported by the National Natural Science Foundation of China (NSFC) under Grant Nos. 12388101, 12125204 and 12502258, and the 111 project of China (No. B17037).}

\backsection[Declaration of interests]
{The authors report no conflict of interest.}

\backsection[Author ORCID]
{
\\
Yi-Zhen Li https://orcid.org/0009-0008-2291-6374
\\
Xin Chen https://orcid.org/0000-0001-9373-9696;
\\
Heng-Dong Xi https://orcid.org/0000-0002-2999-2694.}

%\appendix

\bibliographystyle{jfm}
\bibliography{jfm}

\begin{thebibliography}{62}
\expandafter\ifx\csname natexlab\endcsname\relax\def\natexlab#1{#1}\fi
\def\au#1{#1} \def\ed#1{#1} \def\yr#1{#1}\def\at#1{#1}\def\jt#1{\textit{#1}} \def\bt#1{#1}\def\bvol#1{\textbf{#1}} \def\vol#1{#1} \def\pg#1{#1} \def\publ#1{#1}\def\arxiv#1{#1}\def\org#1{#1}\def\st#1{\textit{#1}}

\bibitem[Ahlers {\em et~al.\/}(2009{\natexlab{{\em a\/}}})Ahlers, Funfschilling \& Bodenschatz]{Ahlers2009NJP}
{\sc \au{Ahlers, G.}, \au{Funfschilling, D.} \& \au{Bodenschatz, E.}} \yr{2009{\natexlab{{\em a\/}}}}  \at{Transitions in heat transport by turbulent convection at {Rayleigh} numbers up to $10^{15}$}.  \jt{New J. Phys.}  \bvol{11},  \pg{123001}.

\bibitem[Ahlers {\em et~al.\/}(2009{\natexlab{{\em b\/}}})Ahlers, Grossmann \& Lohse]{Ahlers2009RMP}
{\sc \au{Ahlers, G.}, \au{Grossmann, S.} \& \au{Lohse, D.}} \yr{2009{\natexlab{{\em b\/}}}}  \at{Heat transfer and large scale dynamics in turbulent {Rayleigh-B\'enard} convection}.  \jt{Rev. Mod. Phys.}  \bvol{81},  \pg{503--537}.

\bibitem[Bouchet {\em et~al.\/}(2019)Bouchet, Rolland \& Simonnet]{bouchet2019rare}
{\sc \au{Bouchet, F.}, \au{Rolland, J.} \& \au{Simonnet, E.}} \yr{2019}  \at{Rare event algorithm links transitions in turbulent flows with activated nucleations}.  \jt{Phys. Rev. Lett.}  \bvol{122}~(7),  \pg{074502}.

\bibitem[Breuer {\em et~al.\/}(2004)Breuer, Wessling, Schmalzl \& Hansen]{Breuer2004PRE}
{\sc \au{Breuer, M.}, \au{Wessling, S.}, \au{Schmalzl, J.} \& \au{Hansen, U.}} \yr{2004}  \at{Effect of inertia in {Rayleigh-B\'enard} convection}.  \jt{Phys. Rev. E}  \bvol{69}~(2),  \pg{026302}.

\bibitem[Brown {\em et~al.\/}(2007)Brown, Funfschilling \& Ahlers]{Brown2007JSM}
{\sc \au{Brown, E.}, \au{Funfschilling, D.} \& \au{Ahlers, G.}} \yr{2007}  \at{Anomalous {Reynolds-number} scaling in turbulent {Rayleigh-B\'enard} convection}.  \jt{J. Stat. Mech.}  \bvol{2007}~(10),  \pg{P10005}.

\bibitem[Chen {\em et~al.\/}(2019)Chen, Huang, Xia \& Xi]{Chen2019JFM}
{\sc \au{Chen, X.}, \au{Huang, S.~D.}, \au{Xia, K.~Q.} \& \au{Xi, H.~D.}} \yr{2019}  \at{Emergence of substructures inside the large-scale circulation induces transition in flow reversals in turbulent thermal convection}.  \jt{J. Fluid Mech.}  \bvol{877},  \pg{R1}.

\bibitem[Chen {\em et~al.\/}(2020)Chen, Wang \& Xi]{Chen2020JFM}
{\sc \au{Chen, X.}, \au{Wang, D.~P.} \& \au{Xi, H.~D.}} \yr{2020}  \at{Reduced flow reversals in turbulent convection in the absence of corner vortices}.  \jt{J. Fluid Mech.}  \bvol{891},  \pg{R5}.

\bibitem[Chong \& Xia(2016)]{Chong2016JFM}
{\sc \au{Chong, K.~L.} \& \au{Xia, K.~Q.}} \yr{2016}  \at{Exploring the severely confined regime in {Rayleigh-B\'enard} convection}.  \jt{J. Fluid Mech.}  \bvol{805},  \pg{R4}.

\bibitem[Dallas {\em et~al.\/}(2020)Dallas, Seshasayanan \& Fauve]{Dallas2020PRF}
{\sc \au{Dallas, V.}, \au{Seshasayanan, K.} \& \au{Fauve, S.}} \yr{2020}  \at{Transitions between turbulent states in a two-dimensional shear flow}.  \jt{Phy. Rev. Fluids}  \bvol{5},  \pg{084610}.

\bibitem[Green {\em et~al.\/}(2020)Green, Vlaykov, Mellado \& Wilczek]{Green2020JFM}
{\sc \au{Green, G.}, \au{Vlaykov, D.~G}, \au{Mellado, J.~P.} \& \au{Wilczek, M.}} \yr{2020}  \at{Resolved energy budget of superstructures in {Rayleigh-B\'enard} convection}.  \jt{J. Fluid Mech.}  \bvol{887},  \pg{A21}.

\bibitem[He {\em et~al.\/}(2021)He, Fang, Gao, Huang \& Bao]{he2021CPB}
{\sc \au{He, J.~C.}, \au{Fang, M.~W.}, \au{Gao, Z.~Y.}, \au{Huang, S.~D.} \& \au{Bao, Y.}} \yr{2021}  \at{Effects of {Prandtl} number in two-dimensional turbulent convection}.  \jt{Chin. Phys. B}  \bvol{30}~(9),  \pg{094701}.

\bibitem[Huang \& Xia(2016)]{Huang2016JFM}
{\sc \au{Huang, S.~D.} \& \au{Xia, K.~Q.}} \yr{2016}  \at{Effects of geometric confinement in quasi-2{D} turbulent {Rayleigh-B\'enard} convection}.  \jt{J. Fluid Mech.}  \bvol{794},  \pg{639--654}.

\bibitem[Huisman {\em et~al.\/}(2014)Huisman, van~der Veen, Sun \& Lohse]{Huisman2014NC}
{\sc \au{Huisman, S.~G.}, \au{van~der Veen, R.~C.}, \au{Sun, C.} \& \au{Lohse, D.}} \yr{2014}  \at{Multiple states in highly turbulent {Taylor-Couette} flow}.  \jt{Nat. Commun.}  \bvol{5},  \pg{3820}.

\bibitem[Kar {\em et~al.\/}(2020)Kar, Kumar, Das \& Lakkaraju]{kar2020thermal}
{\sc \au{Kar, P.~K.}, \au{Kumar, Yada~N.}, \au{Das, P.~K.} \& \au{Lakkaraju, R.}} \yr{2020}  \at{Thermal convection in octagonal-shaped enclosures}.  \jt{Phys. Rev. Fluids}  \bvol{5}~(10),  \pg{103501}.

\bibitem[Krug {\em et~al.\/}(2020)Krug, Lohse \& Stevens]{Krug2020JFM}
{\sc \au{Krug, D.}, \au{Lohse, D.} \& \au{Stevens, R. J. A.~M.}} \yr{2020}  \at{Coherence of temperature and velocity superstructures in turbulent {Rayleigh-B\'enard} flow}.  \jt{J. Fluid Mech.}  \bvol{887},  \pg{A2}.

\bibitem[Lam {\em et~al.\/}(2002)Lam, Shang, Zhou \& Xia]{Lam2002PRE}
{\sc \au{Lam, S.}, \au{Shang, X.~D.}, \au{Zhou, S.~Q.} \& \au{Xia, K.~Q.}} \yr{2002}  \at{{Prandtl} number dependence of the viscous boundary layer and the {Reynolds} numbers in {Rayleigh-B\'enard} convection}.  \jt{Phys. Rev. E}  \bvol{65}~(6),  \pg{066306}.

\bibitem[Lee {\em et~al.\/}(2022)Lee, Lee \& Hwang]{Lee2023design}
{\sc \au{Lee, J.}, \au{Lee, H.} \& \au{Hwang, W.}} \yr{2022}  \at{Design of a compact adjustable laser sheet optical assembly}.  \jt{Meas. Sci. Tech.}  \bvol{34}~(2),  \pg{027003}.

\bibitem[Li {\em et~al.\/}(2021)Li, He, Tian, Hao \& Huang]{Li2021JFM}
{\sc \au{Li, X.~M.}, \au{He, J.~D.}, \au{Tian, Y.}, \au{Hao, P.} \& \au{Huang, S.~D.}} \yr{2021}  \at{Effects of {Prandtl} number in quasi-two-dimensional {Rayleigh-B\'enard} convection}.  \jt{J. Fluid Mech.}  \bvol{915},  \pg{A60}.

\bibitem[Li {\em et~al.\/}(2024{\natexlab{{\em a\/}}})Li, Chen \& Xi]{Li2024JFM}
{\sc \au{Li, Y.~Z.}, \au{Chen, X.} \& \au{Xi, H.~D.}} \yr{2024{\natexlab{{\em a\/}}}}  \at{Enhanced heat transfer and reduced flow reversals in turbulent thermal convection with an obstructed centre}.  \jt{J. Fluid Mech.}  \bvol{981},  \pg{A16}.

\bibitem[Li {\em et~al.\/}(2024{\natexlab{{\em b\/}}})Li, Chen \& Xi]{Li2024POF}
{\sc \au{Li, Y.~Z.}, \au{Chen, X.} \& \au{Xi, H.~D.}} \yr{2024{\natexlab{{\em b\/}}}}  \at{The large-scale circulation and temperature oscillation in turbulent thermal convection in a flattened cylindrical cell of aspect ratio 2}.  \jt{Phys. Fluids}  \bvol{36}~(3),  \pg{035104}.

\bibitem[Li {\em et~al.\/}(2022)Li, Chen, Xu \& Xi]{Li2022JFM}
{\sc \au{Li, Y.~Z.}, \au{Chen, X.}, \au{Xu, A.} \& \au{Xi, H.~D.}} \yr{2022}  \at{Counter-flow orbiting of the vortex centre in turbulent thermal convection}.  \jt{J. Fluid Mech.}  \bvol{935},  \pg{A19}.

\bibitem[Lohse \& Shishkina(2024)]{lohse2024RMP}
{\sc \au{Lohse, D.} \& \au{Shishkina, O.}} \yr{2024}  \at{Ultimate {Rayleigh-B\'enard} turbulence}.  \jt{Rev. Mod. Phys.}  \bvol{96},  \pg{035001}.

\bibitem[Lohse \& Xia(2010)]{Lohse2010ARFM}
{\sc \au{Lohse, D.} \& \au{Xia, K.~Q.}} \yr{2010}  \at{Small-scale properties of turbulent {Rayleigh-B\'enard} convection}.  \jt{Annu. Rev. Fluid Mech.}  \bvol{42},  \pg{335--364}.

\bibitem[Markeviciute \& Kerswell(2021)]{Markeviciute2021JFM}
{\sc \au{Markeviciute, V.~K.} \& \au{Kerswell, R.~R.}} \yr{2021}  \at{Degeneracy of turbulent states in two-dimensional channel flow}.  \jt{J. Fluid Mech.}  \bvol{917},  \pg{A57}.

\bibitem[Moller {\em et~al.\/}(2022)Moller, K{\"a}ufer, Pandey, Schumacher \& Cierpka]{Moller2022JFM}
{\sc \au{Moller, S.}, \au{K{\"a}ufer, T.}, \au{Pandey, A.}, \au{Schumacher, J.} \& \au{Cierpka, C.}} \yr{2022}  \at{Combined particle image velocimetry and thermometry of turbulent superstructures in thermal convection}.  \jt{J. Fluid Mech.}  \bvol{945},  \pg{A22}.

\bibitem[Pandey {\em et~al.\/}(2022)Pandey, Krasnov, Sreenivasan \& Schumacher]{Pandey2022JFM}
{\sc \au{Pandey, A.}, \au{Krasnov, D.}, \au{Sreenivasan, K.~R} \& \au{Schumacher, J.}} \yr{2022}  \at{Convective mesoscale turbulence at very low {Prandtl} numbers}.  \jt{J. Fluid Mech.}  \bvol{948},  \pg{A23}.

\bibitem[Pandey {\em et~al.\/}(2018)Pandey, Scheel \& Schumacher]{Pandey2018NC}
{\sc \au{Pandey, A.}, \au{Scheel, J.~D} \& \au{Schumacher, J.}} \yr{2018}  \at{Turbulent superstructures in {Rayleigh-B\'enard} convection}.  \jt{Nat. Commun.}  \bvol{9}~(1),  \pg{2118}.

\bibitem[Qiu \& Tong(2001)]{Qiu2001PRE}
{\sc \au{Qiu, X.~L.} \& \au{Tong, P.}} \yr{2001}  \at{Large-scale velocity structures in turbulent thermal convection}.  \jt{Phys. Rev. E}  \bvol{64}~(3),  \pg{036304}.

\bibitem[Rahmstorf(2002)]{rahmstorf2002ocean}
{\sc \au{Rahmstorf, S.}} \yr{2002}  \at{Ocean circulation and climate during the past 120,000 years}.  \jt{Nature}  \bvol{419}~(6903),  \pg{207--214}.

\bibitem[Ravelet {\em et~al.\/}(2004)Ravelet, {Mari\'e}, Chiffaudel \& Daviaud]{Ravelet2004PRL}
{\sc \au{Ravelet, F.}, \au{{Mari\'e}, L.}, \au{Chiffaudel, A.} \& \au{Daviaud, F.}} \yr{2004}  \at{Multistability and memory effect in a highly turbulent flow: {Experimental} evidence for a global bifurcation}.  \jt{Phy. Rev. Lett.}  \bvol{93},  \pg{164501}.

\bibitem[Schmeits \& Dijkstra(2001)]{schmeits2001bimodal}
{\sc \au{Schmeits, M.~J.} \& \au{Dijkstra, H.~A.}} \yr{2001}  \at{Bimodal behavior of the kuroshio and the gulf stream}.  \jt{J. Phys. Oceanogr.}  \bvol{31}~(12),  \pg{3435--3456}.

\bibitem[Shishkina(2021)]{Shishkina2021PRF}
{\sc \au{Shishkina, O.}} \yr{2021}  \at{Rayleigh-b{\'e}nard convection: {The} container shape matters}.  \jt{Phys. Rev. Fluids}  \bvol{6}~(9),  \pg{090502}.

\bibitem[Silano {\em et~al.\/}(2010)Silano, Sreenivasan \& Verzicco]{Silano2010JFM}
{\sc \au{Silano, G}, \au{Sreenivasan, KR} \& \au{Verzicco, Roberto}} \yr{2010}  \at{Numerical simulations of {Rayleigh-B\'enard} convection for {Prandtl} numbers between $10^{-1}$ and $10^4$ and {Rayleigh} numbers between $10^5$ and $10^9$}.  \jt{J. Fluid Mech.}  \bvol{662},  \pg{409--446}.

\bibitem[Stevens {\em et~al.\/}(2018)Stevens, Blass, Zhu, Verzicco \& Lohse]{Stevens2018PRF}
{\sc \au{Stevens, R. J. A.~M.}, \au{Blass, A.}, \au{Zhu, X.~J.}, \au{Verzicco, R.} \& \au{Lohse, D.}} \yr{2018}  \at{Turbulent thermal superstructures in {Rayleigh-B\'enard} convection}.  \jt{Phys. Rev. Fluids}  \bvol{3}~(4),  \pg{041501}.

\bibitem[Sugiyama {\em et~al.\/}(2009)Sugiyama, Calzavarini., Grossmann \& Lohse]{Sugiyama2009JFM}
{\sc \au{Sugiyama, K.}, \au{Calzavarini., E.}, \au{Grossmann, S.} \& \au{Lohse, D.}} \yr{2009}  \at{Flow organization in two-dimensional {non-Oberbeck–Boussinesq} {Rayleigh-B\'enard} convection in water}.  \jt{J. Fluid Mech.}  \bvol{637},  \pg{105--135}.

\bibitem[Sugiyama {\em et~al.\/}(2010)Sugiyama, Ni, Stevens, Chan, Zhou, Xi, Sun, Grossmann, Xia \& Lohse]{Sugiyama2010PRL}
{\sc \au{Sugiyama, K.}, \au{Ni, R.}, \au{Stevens, R. J. A.~M.}, \au{Chan, T.~S.}, \au{Zhou, S.~Q.}, \au{Xi, H.~D.}, \au{Sun, C.}, \au{Grossmann, S.}, \au{Xia, K.~Q.} \& \au{Lohse, D.}} \yr{2010}  \at{Flow reversals in thermally driven turbulence}.  \jt{Phys. Rev. Lett.}  \bvol{105}~(3),  \pg{034503}.

\bibitem[{van der Poel} {\em et~al.\/}(2011){van der Poel}, Stevens \& Lohse]{vanderPoel2011PRE}
{\sc \au{{van der Poel}, E.~P.}, \au{Stevens, R. J. A.~M.} \& \au{Lohse, D.}} \yr{2011}  \at{Connecting flow structures and heat flux in turbulent {Rayleigh-B\'enard} convection}.  \jt{Phys. Rev. E}  \bvol{84},  \pg{045303}.

\bibitem[{van der Poel} {\em et~al.\/}(2013){van der Poel}, Stevens \& Lohse]{vanderPoel2013JFM}
{\sc \au{{van der Poel}, E.~P.}, \au{Stevens, R. J. A.~M.} \& \au{Lohse, D.}} \yr{2013}  \at{Comparison between two- and three-dimensional {Rayleigh-B\'enard} convection}.  \jt{J. Fluid Mech.}  \bvol{736},  \pg{177--194}.

\bibitem[{van der Poel} {\em et~al.\/}(2012){van der Poel}, Stevens, Sugiyama \& Lohse]{vanderPoel2012POF}
{\sc \au{{van der Poel}, E.~P.}, \au{Stevens, R. J. A.~M.}, \au{Sugiyama, K.} \& \au{Lohse, D.}} \yr{2012}  \at{Flow states in two-dimensional {Rayleigh-B\'enard} convection as a function of aspect-ratio and {Rayleigh} number}.  \jt{Phys. Fluids}  \bvol{24}~(8),  \pg{085104}.

\bibitem[{van der Veen} {\em et~al.\/}(2016){van der Veen}, Huisman, Dung, Tang, Sun \& Lohse]{vanderVeen2016PRF}
{\sc \au{{van der Veen}, R.~C.}, \au{Huisman, S.~G.}, \au{Dung, O.~Y.}, \au{Tang, H.~L.}, \au{Sun, C.} \& \au{Lohse, D.}} \yr{2016}  \at{Exploring the phase space of multiple states in highly turbulent {Taylor-Couette} flow}.  \jt{Phy. Rev. Fluids}  \bvol{1},  \pg{024401}.

\bibitem[Verzicco \& Camussi(1999)]{verzicco1999JFM}
{\sc \au{Verzicco, R.} \& \au{Camussi, R.}} \yr{1999}  \at{{Prandtl} number effects in convective turbulence}.  \jt{J. Fluid Mech.}  \bvol{383},  \pg{55--73}.

\bibitem[Wang {\em et~al.\/}(2020{\natexlab{{\em a\/}}})Wang, Chong, Stevens, Verzicco \& Lohse]{Wang2020JFM}
{\sc \au{Wang, Q.}, \au{Chong, K.~L.}, \au{Stevens, R. J. A.~M.}, \au{Verzicco, R.} \& \au{Lohse, D.}} \yr{2020{\natexlab{{\em a\/}}}}  \at{From zonal flow to convection rolls in {Rayleigh-B\'enard} convection with free-slip plates}.  \jt{J. Fluid Mech.}  \bvol{905},  \pg{A21}.

\bibitem[Wang {\em et~al.\/}(2020{\natexlab{{\em b\/}}})Wang, Verzicco, Lohse \& Shishkina]{wang2020PRL}
{\sc \au{Wang, Q.}, \au{Verzicco, R.}, \au{Lohse, D.} \& \au{Shishkina, O.}} \yr{2020{\natexlab{{\em b\/}}}}  \at{Multiple states in turbulent large-aspect ratio thermal convection: What determines the number of convection rolls?}  \jt{Phys. Rev. Lett.}  \bvol{125},  \pg{074501}.

\bibitem[Wang {\em et~al.\/}(2018)Wang, Wan \& Sun]{Wang2018PRF}
{\sc \au{Wang, Q.}, \au{Wan, Z.~H.} \& \au{Sun, D.~J.}} \yr{2018}  \at{Multiple states and heat transfer in two-dimensional tilted convection with large aspect ratios}.  \jt{Phys. Rev. Fluids}  \bvol{3},  \pg{113503}.

\bibitem[Wang {\em et~al.\/}(2024)Wang, Chen, Xu \& Xi]{wang2024prandtl}
{\sc \au{Wang, Z.~H.}, \au{Chen, X.}, \au{Xu, A.} \& \au{Xi, H.~D.}} \yr{2024}  \at{Prandtl number dependence of flow topology in quasi-two-dimensional turbulent {Rayleigh-B\'enard} convection}.  \jt{J. Fluid Mech.}  \bvol{991},  \pg{A14}.

\bibitem[Weiss \& Ahlers(2011)]{Weiss2011JFM}
{\sc \au{Weiss, S.} \& \au{Ahlers, G.}} \yr{2011}  \at{Turbulent {Rayleigh-B\'enard} convection in a cylindrical container with aspect ratio {$\Gamma$} = 0.50 and {Prandtl} number {Pr} = 4.38}.  \jt{J. Fluid Mech.}  \bvol{676},  \pg{5--40}.

\bibitem[Xi \& Xia(2008)]{Xi2008POF}
{\sc \au{Xi, H.~D.} \& \au{Xia, K.~Q.}} \yr{2008}  \at{Flow mode transitions in turbulent thermal convection}.  \jt{Phys. Fluids}  \bvol{20},  \pg{055104}.

\bibitem[Xia(2013)]{Xia2013TAML}
{\sc \au{Xia, K.~Q.}} \yr{2013}  \at{Current trends and future directions in turbulent thermal convection}.  \jt{Theor. Appl. Mech. Lett.}  \bvol{3},  \pg{052001}.

\bibitem[Xia {\em et~al.\/}(2002)Xia, Lam \& Zhou]{Xia2002PRL}
{\sc \au{Xia, K.~Q.}, \au{Lam, S.} \& \au{Zhou, S.~Q.}} \yr{2002}  \at{Heat-flux measurement in high-{Prandtl}-number turbulent {Rayleigh-B\'enard} convection}.  \jt{Phys. Rev. Lett.}  \bvol{88}~(6),  \pg{064501}.

\bibitem[Xia {\em et~al.\/}(2008)Xia, Sun \& Cheung]{Xia2008conference}
{\sc \au{Xia, K.~Q.}, \au{Sun, C.} \& \au{Cheung, Y.~H.}} \yr{2008} Large scale velocity structures in turbulent thermal convection with widely varying aspect ratio.  \bt{In {\em 14th International Symposium on Applications of Laser Techniques to Fluid Mechanics\/}}, ,  \vol{vol.~1}.

\bibitem[Xia {\em et~al.\/}(2003)Xia, Sun \& Zhou]{Xia2003PRE}
{\sc \au{Xia, K.~Q.}, \au{Sun, C.} \& \au{Zhou, S.~Q.}} \yr{2003}  \at{Particle image velocimetry measurement of the velocity field in turbulent thermal convection}.  \jt{Phys. Rev. E}  \bvol{68},  \pg{066303}.

\bibitem[Xia {\em et~al.\/}(2018)Xia, Shi, Cai, Wan \& Chen]{xia2018multiple}
{\sc \au{Xia, Z.}, \au{Shi, Y.}, \au{Cai, Q.}, \au{Wan, M.} \& \au{Chen, S.}} \yr{2018}  \at{Multiple states in turbulent plane couette flow with spanwise rotation}.  \jt{J. Fluid Mech.}  \bvol{837},  \pg{477--490}.

\bibitem[Xia {\em et~al.\/}(2017)Xia, Shi, Cai, Wan \& Chen]{Xia2017JFM}
{\sc \au{Xia, Z.~H.}, \au{Shi, Y.~P.}, \au{Cai, Q.~D.}, \au{Wan, M.~P.} \& \au{Chen, S.~Y.}} \yr{2017}  \at{Multiple states in turbulent plane couette flow with spanwise rotation}.  \jt{J. Fluid Mech.}  \bvol{837},  \pg{447--490}.

\bibitem[Xie {\em et~al.\/}(2018)Xie, Ding \& Xia]{Xie2018PRL}
{\sc \au{Xie, Y.~C.}, \au{Ding, G.~Y.} \& \au{Xia, K.~Q.}} \yr{2018}  \at{Flow topology transition via global bifurcation in thermally driven turbulence}.  \jt{Phys. Rev. Lett.}  \bvol{120},  \pg{214501}.

\bibitem[Xie \& Xia(2017)]{Xie2017JFM}
{\sc \au{Xie, Y.~C.} \& \au{Xia, K.~Q.}} \yr{2017}  \at{Turbulent thermal convection over rough plates with varying roughness geometries}.  \jt{J. Fluid Mech.}  \bvol{825},  \pg{573--599}.

\bibitem[Xu {\em et~al.\/}(2023)Xu, Wu \& Xi]{xu2023long}
{\sc \au{Xu, A.}, \au{Wu, H.~L.} \& \au{Xi, H.~D.}} \yr{2023}  \at{Long-distance migration with minimal energy consumption in a thermal turbulent environment}.  \jt{Phys. Rev. Fluids}  \bvol{8}~(2),  \pg{023502}.

\bibitem[Yang {\em et~al.\/}(2020{\natexlab{{\em a\/}}})Yang, Chen, Verzicco \& Lohse]{yang2020multiple}
{\sc \au{Yang, Y.}, \au{Chen, W.}, \au{Verzicco, R.} \& \au{Lohse, D.}} \yr{2020{\natexlab{{\em a\/}}}}  \at{Multiple states and transport properties of double-diffusive convection turbulence}.  \jt{Proc. Natl. Acad. Sci. U.S.A.}  \bvol{117}~(26),  \pg{14676--14681}.

\bibitem[Yang {\em et~al.\/}(2020{\natexlab{{\em b\/}}})Yang, Zhu, Wang, Liu \& Zhou]{Yang2020POF}
{\sc \au{Yang, Y.~H.}, \au{Zhu, X.}, \au{Wang, B.~F.}, \au{Liu, Y.~L.} \& \au{Zhou, Q.}} \yr{2020{\natexlab{{\em b\/}}}}  \at{Experimental investigation of turbulent {Rayleigh-B\'enard} convection of water in a cylindrical cell: The {Prandtl} number effects for {Pr}> 1}.  \jt{Phys. Fluids}  \bvol{32}~(1),  \pg{015101}.

\bibitem[Zhang {\em et~al.\/}(2017)Zhang, Zhou \& Sun]{Zhang2017JFM}
{\sc \au{Zhang, Y.}, \au{Zhou, Q.} \& \au{Sun, C.}} \yr{2017}  \at{Statistics of kinetic and thermal energy dissipation rates in two-dimensional turbulent {Rayleigh-B\'enard} convection}.  \jt{J. Fluid Mech.}  \bvol{814},  \pg{165--184}.

\bibitem[Zhao {\em et~al.\/}(2022)Zhao, Wang, Wu, Chong \& Zhou]{Zhao2022JFM}
{\sc \au{Zhao, C.~B.}, \au{Wang, B.~F.}, \au{Wu, J.~Z.}, \au{Chong, K.~L.} \& \au{Zhou, Q.}} \yr{2022}  \at{Suppression of flow reversals via manipulating corner rolls in plane {Rayleigh-B\'enard} convection}.  \jt{J. Fluid Mech.}  \bvol{946},  \pg{A44}.

\bibitem[Zimmerman {\em et~al.\/}(2011)Zimmerman, Triana \& Lathrop]{zimmerman2011bi}
{\sc \au{Zimmerman, D.~S.}, \au{Triana, S.~A.} \& \au{Lathrop, D.~P.}} \yr{2011}  \at{Bi-stability in turbulent, rotating spherical couette flow}.  \jt{Phys. Fluids}  \bvol{23}~(6).

\bibitem[Zwirner {\em et~al.\/}(2020)Zwirner, Tilgner \& Shishkina]{Zwirner2020PRL}
{\sc \au{Zwirner, L.}, \au{Tilgner, A.} \& \au{Shishkina, O.}} \yr{2020}  \at{Elliptical instability and multiple-roll flow modes of the large-scale circulation in confined turbulent {Rayleigh-B\'enard} convection}.  \jt{Phys. Rev. Lett.}  \bvol{125}~(5),  \pg{054502}.

\end{thebibliography}

\end{document}